\UseRawInputEncoding
\documentclass[
 10pt,
 nofootinbib,
 superscriptaddress,
 twocolumn,
 amsfonts,
 amssymb,
 amsmath,
 aps,
 pra,
 floatfix
]{revtex4-2}
\usepackage{graphicx}
\usepackage{mathtools}      
\usepackage{mathrsfs}
\usepackage{amsthm}
\usepackage{bm}
\usepackage{booktabs}
\usepackage[inline]{enumitem}
\usepackage{csquotes}
\PassOptionsToPackage{colorlinks=true,citecolor=cyan,urlcolor=cyan,linkcolor=cyan,breaklinks=true}{hyperref}
\usepackage{orcidlink}
\usepackage{hyperref}
\usepackage{url}

\newtheoremstyle{break}
  {}{}{\itshape}{}{\bfseries}{.}{\newline}
  {\thmname{#1}\thmnumber{ #2}\thmnote{ (\bfseries #3)}}
\theoremstyle{break}
\newtheorem{theorem}{Theorem}
\newtheorem{lemma}{Lemma}
\newtheorem{definition}{Definition}
\newtheorem{corollary}{Corollary}

\newtheorem{example}{Example}
\providecommand{\openone}{{1\!\!1}}
\newcommand{\expt}[1]{\langle #1 \rangle}
\newcommand{\bra}[1]{\langle #1 |}
\newcommand{\ket}[1]{| #1 \rangle}
\newcommand{\braket}[2]{\langle {#1} | {#2} \rangle}
\newcommand{\ketbra}[2]{| {#1} \rangle\langle {#2} |}
\newcommand{\brakets}[3]{\langle {#1} | {#2} | {#3} \rangle}
\newcommand{\id}{\mathrm{id}}
\newcommand{\Tr}{\mathrm{Tr}}
\newcommand{\Ex}{\mathrm{Ex}}
\newcommand{\Var}{\mathrm{Var}}
\newcommand{\Cov}{\mathrm{Cov}}
\DeclareMathOperator{\rank}{rank}
\newcommand{\e}{\mathrm{e}}
\let\oldd\d
\renewcommand{\d}{\ifmmode\mathrm{d}\else\oldd\fi}
\let\oldi\i
\renewcommand{\i}{\ifmmode\mathrm{i}\else\oldi\fi}
\let\Pr\relax
\newcommand{\Pr}{\mathrm{Pr}}
\let\Re\relax
\DeclareMathOperator{\Re}{Re}
\let\Im\relax
\DeclareMathOperator{\Im}{Im}

\newcommand{\calD}{\mathcal{D}}
\newcommand{\calE}{\mathcal{E}}
\newcommand{\calH}{\mathcal{H}}
\newcommand{\calI}{\mathcal{I}}
\newcommand{\calL}{\mathcal{L}}
\newcommand{\calO}{\mathcal{O}}
\newcommand{\calS}{\mathcal{S}}
\newcommand{\calV}{\mathcal{V}}
\newcommand{\scrC}{\mathscr{C}}
\newcommand{\scrH}{\mathscr{H}}
\newcommand{\scrK}{\mathscr{K}}
\newcommand{\scrM}{\mathscr{M}}
\newcommand{\bbN}{\mathbb{N}}
\newcommand{\bbZ}{\mathbb{Z}}

\newcommand{\bbR}{\mathbb{R}}
\newcommand{\bbC}{\mathbb{C}}
\newcommand{\eq}[1]{\begin{align} #1 \end{align}}
\newcommand{\sa}{\mathrm{sa}}
\def\QED{\mbox{\rule[0pt]{1.5ex}{1.5ex}}}

\def\endproof{\hspace*{\fill}~\QED\par\endtrivlist\unskip}
\newenvironment{proofof}[1]{\vspace*{5mm}\par\noindent
{\bf Proof of #1:\hspace{2mm}}}{\endproof
}

\begin{document}
\title{Quantum statistical functions}
\author{Haruki Emori \orcidlink{0009-0007-2264-9192}}
\email{emori.haruki.i8@elms.hokudai.ac.jp}
\affiliation{Graduate School of Information Science and Technology, Hokkaido University, Kita 14, Nishi 9, Kita-ku, Sapporo, Hokkaido 060-0814, Japan}
\affiliation{RIKEN Center for Interdisciplinary Theoretical and Mathematical Sciences (iTHEMS), RIKEN, 2-1 Hirosawa, Wako, Saitama, 351-0198 Japan}
\date{\today}

\begin{abstract}
Statistical functions such as the moment-generating, characteristic, cumulant-generating, and second characteristic functions are standard tools in classical statistics and probability theory.
They provide a systematic means to analyze the statistical properties of a system and find applications in diverse fields.
While these functions are ubiquitous in classical theory, a quantum counterpart has remained underdeveloped because of the noncommutativity of operators.
The absence of such a framework has obscured the connections between statistical quantities and the nonclassical features of quantum mechanics.
Here, we construct a framework for quantum statistical functions that addresses these limitations and unifies the languages of quantum statistics.
We show that the functions reproduce standard statistical quantities such as expectation values, variance, and covariance upon differentiation.
By extending the framework to include pre- and post-selection, we define conditional functions that generate conditional statistical quantities, including the weak value and the weak variance.
We further show that multivariable functions, defined with specific operator orderings, correspond to  the Kirkwood--Dirac, Margenau--Hill, and Wigner distributions.
By generalizing Bochner's theorem within the theory of compactly supported distributions, we obtain a criterion that separates classical statistics from quantum statistics, linking the failure of positive definiteness of the multivariable function to the emergence of quasiprobability.
As an application, we import the classical method of moments and generalized method of moments into quantum estimation, introducing quantum estimators that exploit the proposed functions.
Our framework reproduces quantum statistical quantities and incorporates the nonclassical features of quasiprobability, providing a basis for further study of quantum statistics.
\end{abstract}
\maketitle

\section{\label{sec:introduction}Introduction}
In classical statistics and probability theory, statistical functions such as the moment-generating function, characteristic function, cumulant-generating function, and second characteristic function play a central role in characterizing the statistical properties of a system under consideration \cite{Feller91,Pistone95,Gibilisco98,Pistone99,Cena07,Pistone13}.
The moment-generating function provides a compact way to encode all the moments of a probability distribution, while the characteristic function, being its Fourier transform, is always well-defined, even when the moments do not exist.
Their logarithmic counterparts, the cumulant-generating and second characteristic functions, are useful for analyzing the independence and additivity of random variables.
These tools are used not only in statistics but also in physics, with applications in statistical mechanics to quantum fluctuation theorems and full counting statistics \cite{Esposito09,Campisi11}, and in quantum field theory to the description of correlation functions \cite{Schwinger51One,Schwinger51Two,Schwinger51PR,Schwinger53,Schenk25}.

Quantum mechanics introduces two features that complicate this classical picture: the noncommutativity of observables and the probabilistic nature of measurement outcomes.
A number of quantum generalizations of individual statistical functions have been developed for specific purposes.
For instance, a moment-generating function defined using quasiprobability distributions has been employed for investigating the fluctuations of non-commuting observables \cite{Hofer17} and the out-of-time-ordered correlators \cite{Yunger-Halpern17,Yunger-Halpern18}.
Similarly, a characteristic function of single observables, defined by the expectation value of a unitary operator, has been used to quantify the nonclassicality of quantum states in bosonic systems \cite{Vogel00,Richter02,Ryl17}.
Furthermore, cumulant-generating functions have seen various specialized adaptations: one, defined as the logarithm of an inner product of parameterized quantum states, connects the parameter space geometry to quantum fluctuations \cite{Hetenyi23}; another, defined via the expectation value of a unitary operator, has been utilized to characterize quantum work statistics \cite{Guarnieri24}.
These constructions, however, each address one statistical function for observables with a fixed ordering, and they do not provide a single object from which the standard moments and cumulants, the multivariable correlation functions associated with different operator orderings, and the conditional moments and cumulants of pre- and post-selected systems are all generated.
In particular, the relation between the standard expectation value and variance on the one hand, and the weak value \cite{Aharonov88} and weak variance \cite{Dressel15,Ogawa21,Matsushita24} based on quasiprobability distributions on the other, has not been organized through a common generating function in which the role of operator ordering is made explicit.
This absence of a unified statistical function---one that explicitly manages operator ordering and seamlessly links standard to conditional moments---constitutes a gap in the current literature.

In this paper, we address this gap by proposing a comprehensive framework for quantum statistical functions.
Specifically, we report the following contributions.

First, we define a set of four single-variable quantum statistical functions: the quantum moment-generating, characteristic, cumulant-generating, and second characteristic functions, each given by an expectation value of an exponential of an observable in the purified state via canonical purification.
We show that they reproduce standard quantum statistical quantities, namely the expectation value, the variance, and the covariance, upon differentiation, and that in the single-variable case they reduce exactly to the classical statistical functions of the projective-measurement distribution.
The multivariable quantum statistical functions reduce to this single-variable case, and hence to the classical functions, whenever the relevant observables commute, so that all nonclassical content is carried by the noncommutativity of the observables.

Second, we introduce a generalized operator-ordering function indexed by a Trotter-type integer $N$ and a normalized weight $w$ on the permutation group, and we show that the multivariable versions defined through it reproduce the Kirkwood--Dirac (KD) \cite{Kirkwood33,Dirac45}, Margenau--Hill (MH) \cite{Margenau61}, and Wigner \cite{Wigner32} distributions for specific choices of $N$ and $w$.
This isolates the statistical role of operator noncommutativity.
By generalizing Bochner's theorem \cite{Bochner55} within the theory of compactly supported distributions, we obtain a criterion that separates classical statistics from quantum statistics: the multivariable quantum characteristic function fails to be positive definite precisely when the associated quasiprobability fails to be a genuine probability measure.
Combining this with the state-dependent notion of commutativity \cite{Ozawa05,Ozawa06,Ozawa14}, we make precise the statement that noncommutativity is necessary but not sufficient for nonclassicality.

Third, we extend the framework from the standard and multivariable quantum statistical functions to \emph{conditional} quantum statistical functions, defined for systems undergoing pre- and post-selection.
The purpose of this extension is to treat conditional quantum statistical properties in general, of which the weak value \cite{Aharonov88} is the first moment and the weak variance \cite{Dressel15,Ogawa21,Matsushita24} the second.
The interpretation of the weak value as a conditional expectation has been advocated and developed in several ways \cite{Steinberg95PRA,Steinberg95PRL,Hofmann11,Dressel14,Spriet25}; our contribution is to place this interpretation on the footing of a conditional moment-generating function, thereby justifying it within the language of quantum statistics, and to show that it reduces to the ordinary conditional expectation when the post-selection commutes with the observable in the relevant state.

Finally, having established a quantum analogue of the classical statistical functions, we import the corresponding inferential machinery.
In classical statistics, the method of moments (MM) and generalized method of moments (GMM) estimate parameters by matching theoretical and empirical moments, and the GMM provides the asymptotically efficient combination of an over-identified set of moment conditions \cite{Hansen82}.
Because our quantum statistical functions generate the theoretical moments and the quantum covariance matrix that plays the role of the optimal GMM weighting, we are able to define a quantum method of moments (QMM) and a quantum generalized method of moments (QGMM) for parameter estimation, which we work out explicitly for the transverse-field Ising model.

The paper is organized as follows.
In Sec.~\ref{sec:preliminary}, we review the foundational concepts: classical statistical functions, the path-integral generating functional, the canonical purification of quantum states, and the matrix geometric mean.
Section~\ref{sec:main} presents the framework: the single-variable, multivariable, and conditional quantum statistical functions, the connection to the KD, MH, and Wigner distributions, the extended Bochner theorem, and the statistical quantities it generates.
Section~\ref{sec:geometry} analyzes the operator-ordering parameter, including convexity properties and a Golden--Thompson comparison.
In Sec.~\ref{sec:app}, we develop the QMM and QGMM.
Section~\ref{sec:alternative} compares our approach with alternatives based on the matrix geometric mean and on information geometry.
We conclude in Sec.~\ref{sec:conclusion}.
Detailed proofs and the connection to discrete Wigner functions are provided in the Appendices.

\section{\label{sec:preliminary}Preliminary}
We first review the essential concepts.
We begin with statistical functions for classical systems, the blueprint for our quantum analogues.
We then discuss the generating functional in quantum field theory.
Next, we describe state purification, the mathematical tool used throughout our definitions.
Finally, we introduce the matrix geometric mean, used in the comparison of Sec.~\ref{sec:alternative}.

\subsection{\label{subsec:csf}Statistical functions for classical systems}
In classical statistics and probability theory, the statistical properties of a random variable $X$ are determined by its probability distribution.
Statistical functions provide an alternative, often more tractable, representation of this information \cite{Feller91,Pistone95,Gibilisco98,Pistone99,Cena07,Pistone13}.

\begin{definition}[Classical moment-generating function]
For a real-valued random variable $X$, the moment-generating function (MGF) $M_{X}(\theta)$ is defined as the expectation value of $\exp(\theta X)$,
\eq{
M_{X}(\theta)\coloneqq\Ex(\e^{\theta X}),
}
where $\theta\in\mathbb{R}$ is a real parameter and $\Ex(\cdots)$ denotes the expectation value.
\end{definition}
The MGF generates the moments $m_{n}$ of the distribution through differentiation, and it equals the (two-sided) Laplace transform of the probability density function of $X$.
A Taylor expansion of the exponential around $\theta=0$ gives
\eq{
M_{X}(\theta)&=\sum_{n=0}^{\infty} \frac{\Ex(X^{n})}{n!}\theta^{n}\nonumber\\
&=\sum_{n=0}^{\infty} \frac{m_{n}}{n!}\theta^{n},
}
provided that the moments exist and the series converges in a neighborhood of $\theta=0$.
The $n$-th moment is then obtained by differentiating the MGF $n$ times with respect to $\theta$ and evaluating at $\theta=0$,
\eq{
m_{n}=\frac{\d^{n}}{\d\theta^{n}}M_{X}(\theta)\Big|_{\theta=0}.
}
The low-order moments are of particular interest: $\Ex(X)$ is the mean, while the central moments, such as the variance $\Var(X)\coloneqq\Ex\{[X-\Ex(X)]^{2}\}$, the skewness $\Ex\{[X-\Ex(X)]^{3}\}/\Var(X)^{3/2}$, and the kurtosis $\Ex\{[X-\Ex(X)]^{4}\}/\Var(X)^{2}$, describe the shape of the distribution.
A limitation of the MGF is that it is not guaranteed to exist for all values of $\theta$, since the integral or sum defining the expectation value may diverge.
This convergence issue is avoided by the characteristic function.

\begin{definition}[Classical characteristic function]
The characteristic function (CF) $C_{X}(\theta)$ of a random variable $X$ is defined as the expectation value of $\exp(\i\theta X)$,
\eq{
C_{X}(\theta)\coloneqq\Ex(\e^{\i\theta X}),
}
where $\i$ is the imaginary unit.
\end{definition}
Since $|\exp(\i\theta X)| = 1$, the expectation value always converges, so the CF exists for all real $\theta$ and for any random variable.
It is related to the MGF by analytic continuation to an imaginary argument,
\eq{
C_{X}(\theta)=M_{X}(\i\theta),
}
and equals the Fourier transform of the probability distribution of $X$.
The moments are obtained as
\eq{
m_{n}=\i^{-n}\frac{\d^{n}}{\d\theta^{n}}C_{X}(\theta)\Big|_{\theta=0}.
}
Cumulants provide complementary information about a distribution, in particular its additivity for independent random variables.
Cumulants are generated by the logarithm of the MGF.

\begin{definition}[Classical cumulant-generating function]
The cumulant-generating function (CGF) $K_{X}(\theta)$ is defined as
\eq{
K_{X}(\theta)\coloneqq\log M_{X}(\theta).
}
\end{definition}
The Taylor expansion of the CGF yields the cumulants $k_{n}$ as coefficients,
\eq{
K_{X}(\theta)=\sum_{n=1}^{\infty}\frac{k_{n}}{n!}\theta^{n},
}
so the $n$-th cumulant is
\eq{
k_{n}=\frac{\d^{n}}{\d\theta^{n}}K_{X}(\theta)\Big|_{\theta=0}.
}
The first cumulant $k_{1}$ is the mean and the second cumulant $k_{2}$ is the variance.
Higher-order cumulants are related to, but distinct from, the higher-order central moments.
The logarithm of the CF generates the same cumulants.

\begin{definition}[Classical second characteristic function]
The second characteristic function $H_{X}(\theta)$ is defined as
\eq{
H_{X}(\theta)\coloneqq\log C_{X}(\theta).
}
\end{definition}
The cumulants are obtained by differentiation,
\eq{
k_{n}=\i^{-n}\frac{\d^{n}}{\d\theta^{n}}H_{X}(\theta)\Big|_{\theta=0}.
}
These four functions form the toolkit for the analysis of classical statistical properties that we extend to quantum systems.

\subsection{\label{subsec:pi}Path integral and generating functional}
In quantum field theory, the generating function becomes a generating functional from which the physical quantities are derived.
The path-integral formalism provides the natural language for this construction \cite{Feynman48,Dyson49,Feynman49}.
The central quantity is the transition amplitude, or propagator, expressed as a sum over all paths of a system.
For a particle moving from $(x_i, t_i)$ to $(x_f, t_f)$, it is given by
\eq{
K(x_{f},t_{f};x_{i},t_{i})=\int\calD[x(t)]\e^{\frac{\i}{\hbar}S[x(t)]},
}
where
\eq{
S[x(t)] = \int_{t_i}^{t_f}\calL(x(t),\dot{x}(t),t)\d t
}
is the classical action and the integral is a functional integral over all paths $x(t)$ connecting the endpoints.
This idea underlies the generating functional $Z[J]$ for a quantum field $\phi$ \cite{Schwinger51One,Schwinger51Two}, coupled to an external classical source field $J(x)$ \cite{Schwinger51PR,Schwinger53}.
The functional $Z[J]$ is then defined as the vacuum-to-vacuum transition amplitude in the presence of this source:
\eq{
Z[J]\coloneqq\int \mathcal{D}[\phi] \, \e^{\i \left( S[\phi] + \expt{J, \phi} \right)},\label{eq:partition_func}
}
where $\expt{J, \phi} = \int \d^{4} x J(x)\phi(x)$ is the coupling of the field to the source.
The functional $Z[J]$ encodes the dynamics and correlations of the field $\phi$.
The vacuum expectation values of time-ordered products of field operators, the $n$-point correlation functions, are generated by taking functional derivatives of $Z[J]$ with respect to $J(x)$,
\eq{
\bra{0}T\{\phi(x_1)\cdots\phi(x_n)\}\ket{0} = \frac{1}{\i^{n}Z[0]}\left. \frac{\delta^n Z[J]}{\delta J(x_1)\cdots\delta J(x_n)}\right|_{J=0}.
}
This identifies $Z[J]$ as the field-theoretic counterpart of the characteristic function.
The logarithm $W[J] = -\i \log Z[J]$ is the generating functional for connected Green's functions, analogous to the cumulant-generating function.
The observation that a single object generates all relevant statistical correlations motivates the construction below.

\subsection{\label{subsec:purification}Purification}
The state of a quantum system is described by a density operator $\rho$, a positive semidefinite operator with unit trace.
If $\rho$ is a pure state, it can be written as a projector $\rho = \ketbra{\psi}{\psi}$.
If the system is in a mixed state, $\rho$ cannot be written this way.
Purification represents a mixed state by embedding it into a larger system that is in a pure state.

\subsubsection{\label{subsubsec:conv}Conventional purification}
Any mixed state $\rho$ acting on a Hilbert space $\mathcal{H}$ can be represented as a pure state $\ket{\Psi}$ in an extended Hilbert space $\mathcal{H} \otimes \mathcal{H}'$, where $\mathcal{H}'$ is an ancillary Hilbert space.
This pure state $\ket{\Psi}$ is a purification of $\rho$.
The original density matrix is recovered by tracing out the ancillary system,
\eq{
\rho=\Tr_{\calH'}(\ketbra{\Psi}{\Psi}).
}
If the spectral decomposition of $\rho$ is $\rho = \sum_{i} \lambda_i \ketbra{\alpha_i}{\alpha_i}$, a purification can be constructed using the Schmidt decomposition.
For any orthonormal set $\{\ket{\beta_i}\}$ in an ancillary Hilbert space $\mathcal{H}'$ with $\dim(\mathcal{H}') \ge \rank(\rho)$, the state
\eq{
\ket{\Psi}=\sum_{i}\sqrt{\lambda_{i}}\ket{\alpha_{i}}\otimes\ket{\beta_{i}}
}
is a purification of $\rho$.
This construction is not unique, as it depends on the choice of the set $\{\ket{\beta_i}\}$ in $\mathcal{H}'$.

\subsubsection{\label{subsubsec:cano}Canonical purification}
A statistical function should depend on the physical data $(\rho,A)$ alone, not on the gauge freedom of the purification.
We therefore fix this freedom from the outset by adopting a canonical choice.
We take the ancillary space $\mathcal{H}'$ to be the dual space $\calH^{*}$ of the original Hilbert space.
This space consists of all bra vectors corresponding to the ket vectors in $\calH$.
The inner product on $\calH^{*}$ is defined by
\eq{
(\bra{\zeta},\bra{\eta})=\braket{\eta}{\zeta}
}
for all $\ket{\zeta}, \ket{\eta}\in\calH$.
For any operator $A$ on $\calH$ and all $\ket{\eta}\in\calH$, we define the operator $A^{*}$ on $\calH^{*}$ by
\eq{
A^{*}\bra{\eta}=\bra{\eta}A,
}
so that
\eq{
(\bra{\zeta},A^{*}\bra{\eta})=\brakets{\eta}{A}{\zeta}
}
for all $\ket{\zeta},\ket{\eta}\in\calH$.
We now set $\calH'=\calH^{*}$, which leads to the canonical purification \cite{VonNeumann55,Ozawa14}.
\begin{definition}[Canonical purification]
For a state $\rho$ with spectral decomposition $\rho=\sum_{i}\lambda_{i}\ketbra{\alpha_{i}}{\alpha_{i}}$, its canonical purification is the pure state $\ket{\Psi}\in\calH\otimes\calH^{*}$ defined by
\eq{
\ket{\Psi}=\sum_{i}\sqrt{\lambda_{i}}\ket{\alpha_{i}}\otimes\bra{\alpha_{i}}.
}
\end{definition}
This definition is unique up to a choice of basis for degenerate eigenvalues.
The tensor-product space $\calH\otimes\calH^{*}$ is isomorphic to the space $\calL(\calH)$ of Hilbert--Schmidt operators through the correspondence between $\ket{\alpha}\otimes\bra{\alpha}\in\calH\otimes\calH^{*}$ and the operator $T_{\ket{\alpha}\otimes\bra{\alpha}}\coloneqq\ketbra{\alpha}{\alpha}\in\calL(\calH)$, where $T_{\ket{\alpha}\otimes\bra{\alpha}}\ket{\psi}=\braket{\alpha}{\psi}\ket{\alpha}$ for all $\ket{\psi}\in\calH$.
Under this isomorphism, the canonical purification $\ket{\Psi}\in\calH\otimes\calH^{*}$ of a density operator $\rho$ corresponds to $\sqrt{\rho}\in\calL(\calH)$.
Thus any proof using the operator $\sqrt{\rho}\in\calL(\calH)$ can be transferred to a proof using the canonical purification $\ket{\Psi}\in\calH\otimes\calH^{*}$ of $\rho$.
For any operator $X \in \calL(\calH)$, its action on the purified state in the extended Hilbert space $\calH \otimes \calH^{*}$ is defined by the extension $X \otimes \openone_{\calH^{*}}$, where $\openone_{\calH^{*}}$ is the identity operator on $\calH^{*}$.
With this extension, expectation values computed in the purified-state picture coincide with $\Tr(X \rho)$ in the mixed-state picture, since
\eq{
\bra{\Psi}(X \otimes \openone_{\calH^{*}})\ket{\Psi} &= \sum_{i}\lambda_i \bra{\alpha_i}X\ket{\alpha_i}\nonumber\\
&= \Tr(X\rho),\label{eq:purified_trace}
}
which is manifestly independent of the purification gauge.
All our quantum statistical functions are defined as expectation values, or inner products, with respect to this canonical purification; this gauge invariance is a property we require of the construction rather than a result, and the canonical purification is the device that secures it.

\subsection{\label{subsec:mgm}Matrix geometric mean and amplitudes}
We here define the matrix geometric mean and describe its structural properties.
For two positive definite operators $A, B > 0$, the arithmetic mean is not the only way to define a center.
The geometric mean $A \# B$ \cite{Pusz75,Kubo80} is defined as
\eq{
A \# B \coloneqq A^{1/2} \left( A^{-1/2} B A^{-1/2} \right)^{1/2} A^{1/2}. \label{eq:def_MGM}
}
This quantity is the midpoint of the geodesic connecting $A$ and $B$ on the Riemannian manifold of positive definite matrices equipped with the trace metric $\d s^{2} = \Tr[(A^{-1}\d A)^{2}]$.
The geodesic $\gamma(u) = A \#_{u} B = A^{1/2} (A^{-1/2} B A^{-1/2})^{u} A^{1/2}$ for $u \in [0,1]$ interpolates between $A$ and $B$, with the geometric mean at $u=1/2$.

\subsubsection{\label{subsubsec:vari}Variational characterization}
The geometric mean admits a variational characterization.
The geometric mean $A \# B$ is characterized as the solution to a maximization problem within the cone of positive operators: $A \# B$ is the largest positive operator $X$ satisfying a block-matrix positivity condition,
\eq{
A \# B = \max \left\{ X \ge 0 \;\middle|\; \begin{pmatrix} A & X \\ X & B \end{pmatrix} \ge 0 \right\}.
}
This connects to the quantum fidelity $F(A, B) = \Tr|\sqrt{A}\sqrt{B}|$, which can be expressed by a similar block-matrix condition, but with a maximization over the trace of a general contraction $X$,
\eq{
F(A, B) = \sup \left\{ |\Tr(X)| \;\middle|\; \begin{pmatrix} A & X \\ X^\dagger & B \end{pmatrix} \ge 0 \right\}.
}
Comparing the two, the geometric mean $A \# B$ captures the structural overlap between $A$ and $B$ without the unitary rotation that maximizes the fidelity.
Consequently $\Tr(A \# B) \le F(A, B)$, with equality if and only if $[A, B] = 0$.

\subsubsection{\label{subsubsec:amp}Relation to amplitudes and to the purification}
An amplitude $W$ of a density operator $\rho$ is any operator satisfying $\rho=WW^{\dagger}$ \cite{Uhlmann76}.
In the Hilbert--Schmidt picture, an amplitude is exactly a representative of the purification: the canonical amplitude $W_\rho=\sqrt{\rho}$ corresponds to the canonical purification $\ket{\Psi}$ of Sec.~\ref{subsubsec:cano}, while a general amplitude $W_\rho=\sqrt{\rho}\,U$, where $U$ is a unitary operator, corresponds to the gauge transformation $\ket{\Psi}\mapsto(\openone\otimes U^{T*})\ket{\Psi}$ of the ancilla.
For two states $\rho$ and $\sigma$ with amplitudes $W_{\rho}$ and $W_{\sigma}$, the amplitudes are parallel if $W_{\rho}^{\dagger} W_{\sigma} \ge 0$, which fixes this gauge freedom inherent in the purification of mixed states.
For parallel amplitudes, the geometric mean appears as the transformation ratio,
\eq{
W_{\sigma} W^{-1}_{\rho} = \sigma \# \rho^{-1}.
}
The geometric mean thus acts as the operator that maps one amplitude to another along the parallel-transport alignment; we use this in Sec.~\ref{subsec:gmb}.

\section{\label{sec:main}Main results}
We now construct statistical functions for quantum systems.
We write $\calL(\calH)$ for the bounded operators on $\calH$ and $\calL_{\sa}(\calH)\subset\calL(\calH)$ for the bounded self-adjoint operators (observables).

\subsection{\label{subsec:qsf}Statistical functions for quantum systems}
We define quantum statistical functions as the inner product between a purified state $\ket{\Psi}$ of interest $\rho$ and the purified state obtained by applying an operator-valued function; equivalently, as the expectation value of that function in $\ket{\Psi}$.

\subsubsection{\label{subsubsec:single}Single-variable case}
We first treat the single-variable case.

\begin{definition}[Quantum moment-generating function]
For a purified state $\ket{\Psi} \in \calH \otimes \calH^*$ of $\rho$ and an observable $A \in \calL_{\text{sa}}(\calH)$, the quantum moment-generating function (QMGF) is defined by
\eq{
\scrM_{A}(\theta,\rho)\coloneqq\bra{\Psi}(\e^{\theta A} \otimes \openone_{\calH^{*}})\ket{\Psi},\label{eq:qmgf}
}
where $\theta \in \mathbb{R}$.
\end{definition}
For a self-adjoint operator $A$, the operator $\exp(\theta A)$ is positive definite but not unitary for $\theta \ne 0$; it can be read as an imaginary-time evolution operator.
This function is the quantum counterpart of the classical MGF.
Unlike the multivariable cases discussed below, the single-variable QMGF is real-valued and strictly positive for any bounded self-adjoint operator $A$ and any full-rank state $\rho$.
This follows from the identity $\bra{\Psi}[\exp(\theta A) \otimes \openone_{\calH^{*}}]\ket{\Psi} = \Tr[\exp(\theta A)\rho]$ established in Eq.~\eqref{eq:purified_trace}: since $\exp(\theta A)$ and $\rho$ are both positive definite for $\theta \in \bbR$, the trace of their product is real and positive.
This guarantees that the quantum cumulant-generating function below is well-defined.

\begin{definition}[Quantum characteristic function]
For a purified state $\ket{\Psi} \in \calH \otimes \calH^*$ of $\rho$ and an observable $A \in \calL_{\text{sa}}(\calH)$, the quantum characteristic function (QCF) is defined by
\eq{
\scrC_{A}(\theta,\rho)\coloneqq\bra{\Psi}(\e^{\i\theta A}\otimes\openone_{\calH^{*}})\ket{\Psi},\label{eq:qcf}
}
where $\exp(\i\theta A)$ is the unitary operator generated by $A$.
\end{definition}
This function is well-defined for all real $\theta$.
As in the classical case, it is related to the QMGF by an imaginary argument,
\eq{
\scrC_{A}(\theta,\rho)= \scrM_{A}(\i\theta,\rho),
}
which mirrors the Wick rotation \cite{Wick54}.

\begin{definition}[Quantum cumulant-generating function]
For a purified state $\ket{\Psi} \in \calH \otimes \calH^*$ of $\rho$ and an observable $A \in \calL_{\text{sa}}(\calH)$, the quantum cumulant-generating function (QCGF) is defined as the logarithm of the quantum moment-generating function,
\eq{
\scrK_{A}(\theta,\rho)\coloneqq\log\scrM_{A}(\theta,\rho).\label{eq:qcgf}
}
\end{definition}
A logarithmic version of the quantum second characteristic function can be introduced analogously.

\begin{definition}[Quantum second characteristic function]
For a purified state $\ket{\Psi} \in \calH \otimes \calH^*$ of $\rho$ and an observable $A \in \calL_{\text{sa}}(\calH)$, the quantum second characteristic function (QSCF) is defined as the logarithm of the quantum characteristic function,
\eq{
\scrH_{A}(\theta,\rho)\coloneqq\log\scrC_{A}(\theta,\rho).\label{eq:qscf}
}
\end{definition}
These single-variable functions reduce to their classical counterparts.
Let $A=\sum_{a} a\,P_{A}(a)$ be the spectral decomposition of $A$, with $P_{A}(a)$ the projection onto the eigenspace of eigenvalue $a$.
Using Eq.~\eqref{eq:purified_trace},
\eq{
\scrM_{A}(\theta,\rho)&=\Tr(\e^{\theta A}\rho)\nonumber\\
&=\sum_{a}\e^{\theta a}\,\Pr\{A=a\Vert\rho\},\label{eq:qmgf_classical}
}
where $\Pr\{A=a\Vert\rho\}\coloneqq\Tr[P_{A}(a)\rho]\ge 0$ is the probability of obtaining the outcome $a$ axiomatized by the Born statistical formula, and $\sum_{a}\,\Pr\{A=a\Vert\rho\}=1$.
Hence the single-variable QMGF is exactly the classical MGF of the random variable defined by a projective measurement of $A$ in the state $\rho$; the QCF, QCGF, and QSCF reduce to the classical CF, CGF, and second characteristic function of the same distribution.
The nonclassical content of our framework therefore appears only in the multivariable and conditional cases, where the noncommutativity of the operators involved cannot be removed by a single spectral decomposition.

\subsubsection{\label{subsubsec:multi}Multivariable case}
Extending the single-variable definitions to several variables is nontrivial because operators do not commute.
For classical random variables, the multivariable statistical functions are unambiguously defined through $\exp(\sum_{j}\theta_{j}X_{j})$.
In the quantum case, $\exp(\sum_{j}\theta_{j}A_{j})\ne \prod_{j}\exp(\theta_{j}A_{j})$ in general, and the order of operators in such a product matters; the analysis of experimentally measured statistics, in particular full counting statistics, requires a framework that handles this ordering explicitly.
To this end, we introduce a generalized operator-ordering function $f_{\bm{A}}^{(N, w)}(\bm{\theta})$, which provides a unified description of the quasiprobability distributions \cite{Ferrie11,Lee17,Lee18,Arvidsson-Shukur21,Langrenez24,Arvidsson-Shukur24} (also called weak joint distributions \cite{Ozawa05,Ozawa11}) used in quantum mechanics.

\begin{definition}[Generalized operator-ordering function]
Let $\bm{A} = \{A_1, \dots, A_n\}\subset\calL_{\sa}(\calH)$ be a set of observables and $\bm{\theta} = \{\theta_1, \dots, \theta_n\} \in \mathbb{R}^n$ be the associated parameters.
Let $N\in\bbN$ be a positive integer and $w: S_n \to \mathbb{C}$ be a normalized weight function on the permutation group $S_n$, with $\sum_{\sigma \in S_n} w(\sigma) = 1$.
The generalized operator-ordering function is defined by
\eq{
f_{\bm{A}}^{(N, w)}(\bm{\theta}) \coloneqq \left[ \sum_{\sigma \in S_n} w(\sigma) \left( \prod_{j=1}^{n} \e^{\frac{\theta_{\sigma(j)}}{N} A_{\sigma(j)}} \right) \right]^N. \label{eq:gen_op_func}
}
\end{definition}

We remark that each factor $\exp[\theta_{\sigma(j)}A_{\sigma(j)}/N]$ is a bounded operator with $\|\exp(\theta A/N)\|\le\exp(|\theta|\,\|A\|/N)$ (Appendix~\ref{sec:extended_bochner}).
The product over $j$, the finite weighted sum over $S_n$, and the $N$-th power are therefore well-defined bounded operators for any bounded $\bm{A}$ and any complex weight $w$; the normalization $\sum_\sigma w(\sigma)=1$ gives $f_{\bm{A}}^{(N,w)}(\bm{0})=\openone$.
Equation~\eqref{eq:gen_op_func} generalizes the scalar identity $\exp(X)=\lim_{N\to\infty}(1+X/N)^{N}$ to the noncommutative setting, interpolating between products of exponentials ($N=1$) and the exponential of the sum ($N\to\infty$).

\begin{definition}[Multivariable quantum moment-generating function]
For a set $\bm{A} = \{A_1, \dots, A_n\}\subset\calL_{\text{sa}}(\calH)$ of observables and real parameters $\bm{\theta} = \{\theta_1, \dots, \theta_n\} \in \mathbb{R}^n$, the multivariable QMGF in a state $\rho$ with canonical purification $\ket{\Psi}$ is defined by
\eq{
\scrM_{\bm{A}}(\bm{\theta},\rho)\coloneqq\bra{\Psi}(f_{\bm{A}}^{(N, w)}(\bm{\theta})\otimes\openone_{\calH^{*}})\ket{\Psi}.\label{eq:mqmgf}
}
\end{definition}
For specific choices of $N$ and $w$, this function reproduces three quasiprobability distributions.

\paragraph{Kirkwood--Dirac (KD) distribution:}
Set $N=1$ and choose the deterministic weight concentrated on the identity permutation, $w(\sigma) = \delta_{\sigma, \id}$.
The function reduces to the sequentially ordered product,
\eq{
f_{\bm{A}}^{\text{KD}}(\bm{\theta}) &= \e^{\theta_{1} A_{1}} \e^{\theta_{2} A_{2}} \cdots \e^{\theta_{n} A_{n}} \nonumber\\
&= \prod_{j=1}^{n} \e^{\theta_{j} A_{j}}.\label{eq:kd_dist}
}
The corresponding QMGF,
\eq{
\scrM^{\text{KD}}_{\bm{A}}(\bm{\theta},\rho)&=\bra{\Psi}(f^{\text{KD}}_{\bm{A}}(\bm{\theta})\otimes\openone_{\calH^{*}})\ket{\Psi}\nonumber\\
&=\Tr\left(\prod^{n}_{j=1}\e^{\theta_{j}A_{j}}\rho\right),\label{eq:kd-qmgf}
}
generates the KD distribution \cite{Kirkwood33, Dirac45}.

\paragraph{Margenau--Hill (MH) distribution:}
Set $N=1$ and choose the symmetrized weight $w(\id) = w(\tau) = 1/2$, where $\tau$ is the transposition $(1, 2)$ (for $n=2$; for general $n$, $\tau$ is the order-reversing permutation).
The function reduces to a convex combination of the forward and reverse orderings,
\eq{
f^{\text{MH}}_{\bm{A}}(\bm{\theta})=\frac{1}{2}\left(\prod^{n}_{j=1}\e^{\theta_{j}A_{j}}+\prod^{n-1}_{j=0}\e^{\theta_{n-j}A_{n-j}}\right).\label{eq:mh_dist}
}
The corresponding QMGF,
\eq{
\scrM^{\text{MH}}_{\bm{A}}(\bm{\theta},\rho)&=\bra{\Psi}(f^{\text{MH}}_{\bm{A}}(\bm{\theta})\otimes\openone_{\calH^{*}})\ket{\Psi}\nonumber\\
&=\frac{1}{2}\Tr\left(\prod^{n}_{j=1}\e^{\theta_{j}A_{j}}\rho\right)+\frac{1}{2}\Tr\left(\prod^{n-1}_{j=0}\e^{\theta_{n-j}A_{n-j}}\rho\right),\label{eq:mh-qmgf}
}
generates the real part of the KD distribution, namely the MH distribution \cite{Margenau61}.

\paragraph{Wigner distribution:}
Take the limit $N \to \infty$.
By the Lie--Trotter product formula, the ordering of the infinitesimal generators is irrelevant at first order in $1/N$,
\eq{
\prod_{j=1}^{n} \e^{\frac{\theta_{\sigma(j)}}{N} A_{\sigma(j)}} = \openone + \frac{1}{N} \sum_{j=1}^n \theta_j A_j + O(1/N^2).
}
Substituting into Eq.~\eqref{eq:gen_op_func} and using $\sum_\sigma w(\sigma)=1$, the sum over permutations yields the identity plus the average sum, independent of $\sigma$ at this order, so that
\eq{
f_{\bm{A}}^{\text{W}}(\bm{\theta}) &= \lim_{N \to \infty} \left( \openone + \frac{1}{N} \sum_{j=1}^n \theta_j A_j \right)^N \nonumber\\
&= \exp\left( \sum_{j=1}^n \theta_j A_j \right).
}
The corresponding QMGF is
\eq{
\scrM^{\text{W}}_{\bm{A}}(\bm{\theta},\rho)&=\bra{\Psi}(f^{\text{W}}_{\bm{A}}(\bm{\theta})\otimes\openone_{\calH^{*}})\ket{\Psi}\nonumber\\
&=\Tr\left(\e^{\sum_{j=1}^n \theta_j A_j}\rho\right).\label{eq:w-qmgf}
}
This is the QMGF corresponding to the Weyl symmetric ordering.
The connection to the Wigner distribution \cite{Wigner32} appears when one passes from the moment-generating domain to the characteristic-function domain.
By replacing the real parameters $\bm{\theta}$ with imaginary arguments $\i\bm{\theta}$, we see that the operator $D_{\bm{A}}(\bm{\theta}) \dots$ becomes the unitary displacement operator on phase space (assuming $\bm{A}$ are canonical observables such as position and momentum).
The expectation value $\bra{\Psi} (D_{\bm{A}}(\bm{\theta})\otimes\openone_{\calH^{*}}) \ket{\Psi}$ is then the Wigner-type quantum characteristic function $\scrC_{\bm{A}}^{\text{W}}(\bm{\theta}, \rho)$, and the Wigner distribution is recovered by the symplectic Fourier transform of this characteristic function.
In the limit $N \to \infty$, our framework thus defines the generator in the dual domain of the Wigner function.
The mathematical subtleties of this correspondence, in particular the stratification of phase space for discrete finite-dimensional systems \cite{Wootters87,Gross06,DeBrota20}, are treated in Appendix~\ref{sec:dwf}.

To make the connection explicit, consider the two-variable case for a state $\rho$ with $f^{\text{KD}}_{A,B}(\theta_{1},\theta_{2})=\exp(\theta_{2} B)\exp(\theta_{1}A)$.
The KD-type QMGF is the generating function of the KD distribution,
\eq{
\scrM^{\text{KD}}_{A,B}(\theta_{1},\theta_{2})&=\bra{\Psi}(f^{\text{KD}}_{A,B}(\theta_{1},\theta_{2})\otimes\openone_{\calH^{*}})\ket{\Psi}\nonumber\\
&=\sum_{a,b}\e^{\theta_{1}a+\theta_{2}b}\,\Tr[P_{B}(b)P_{A}(a)\rho]\nonumber\\
&=\sum_{a,b}\e^{\theta_{1}a+\theta_{2}b}\,\Pr^{\text{KD}}\{A=a,B=b\Vert\rho\},
}
where $P_{A}(a)$ and $P_{B}(b)$ are spectral projections onto the eigenspaces of $A$ and $B$, and
\eq{
\Pr^{\text{KD}}\{A=a,B=b\Vert\rho\}=\Tr\left[P_{B}(b)P_{A}(a)\rho\right]
}
is the KD distribution for obtaining the outcome $A=a$ and then $B=b$.
This distribution permits the calculation of weak values \cite{Aharonov88,Dressel14}.
The conditional quasiprobability of obtaining an outcome $A=a$ given a subsequent outcome $B=b$ in a pure state $\psi=\ketbra{\psi}{\psi}$ is
\eq{
\Pr^{\text{KD}}\{A=a|B=b\Vert\psi\}&=\frac{\Pr^{\text{KD}}\{A=a,B=b\Vert\psi\}}{\Pr\{B=b\Vert\psi\}}\nonumber\\
&=\frac{\braket{b}{a}\braket{a}{\psi}}{\braket{b}{\psi}}.\label{eq:con-quasiprob}
}
Using Eq.~\eqref{eq:con-quasiprob}, the conditional expectation of $A$ given post-selection on $B=b$ is
\eq{
\Ex^{\text{KD}}_{\psi}(A|B=b)&=\sum_{a} a\,\Pr^{\text{KD}}\{A=a|B=b\Vert\psi\}\nonumber\\
&=\frac{\bra{b}A\ket{\psi}}{\braket{b}{\psi}},
}
which is the weak value of $A$ for a system pre-selected in $\ket{\psi}$ and post-selected in $\ket{b}$ \cite{Steinberg95PRA,Steinberg95PRL}.
The nonclassical values of the KD distribution are signatures of the noncommutativity of $A$ and $B$, and they have been related to quantum contextuality and to a resource for advantage in metrology \cite{Jordan14,Lostaglio20,Arvidsson-Shukur20}.

We next introduce a variant of the generalized operator-ordering function for the unitary operators.
\begin{definition}[Generalized unitary operator-ordering function]
Let $\bm{A} = \{A_1, \dots, A_n\}\subset\calL_{\sa}(\calH)$ be a set of self-adjoint observables and $\bm{\theta} = \{\theta_1, \dots, \theta_n\} \in \mathbb{R}^n$ be the associated real parameters.
Let $N \in \mathbb{N}$ be a positive integer and $w: S_n \to \mathbb{C}$ be a normalized weight function on $S_n$, with $\sum_{\sigma \in S_n} w(\sigma) = 1$.
The generalized unitary operator-ordering function is defined by
\eq{
\tilde{f}_{\bm{A}}^{(N, w)}(\bm{\theta}) \coloneqq \left[ \sum_{\sigma \in S_n} w(\sigma) \left( \prod_{j=1}^{n} \e^{\frac{\i\theta_{\sigma(j)}}{N} A_{\sigma(j)}} \right) \right]^N.\label{eq:gen_uni_func}
}
\end{definition}
In this construction, the real parameters are replaced by imaginary arguments, ensuring that each exponential factor is a unitary operator with a spectral norm of exactly one,  $\|\exp(\i\theta A/N)\|=1$.
Consequently, the triangle inequality guarantees that the generalized unitary operator-ordering function is uniformly bounded; its norm is strictly bounded by the sum of the absolute weights $\|\tilde f_{\bm{A}}^{(N,w)}(\bm\theta)\|\le\sum_\sigma|w(\sigma)|$.
Furthermore, evaluating this function at the origin trivially yields the identity operator $\tilde f_{\bm{A}}^{(N,w)}(\bm0)=\openone$.
The multivariable QCF is defined using this function.
\begin{definition}[Multivariable quantum characteristic function]
For a set $\bm{A} = \{A_1, \dots, A_n\} \subset \calL_{\text{sa}}(\calH)$ of observables and real parameters $\bm{\theta} = \{\theta_1, \dots, \theta_n\} \in \mathbb{R}^n$, the multivariable QCF in a state $\rho$ with canonical purification $\ket{\Psi}$ is defined by
\eq{
\scrC_{\bm{A}}(\bm{\theta},\rho)\coloneqq\bra{\Psi}(\tilde{f}_{\bm{A}}^{(N, w)}(\bm{\theta})\otimes\openone_{\calH^{*}})\ket{\Psi}.\label{eq:mqcf}
}
\end{definition}
The multivariable QCF characterizes the nonclassicality of the state through the structure of the underlying quasiprobability.
We formalize this as an extension of Bochner's theorem \cite{Bochner55} within the theory of compactly supported distributions \cite{Reed80,Hormander03}.
We first recall the two classical inputs and fix the function spaces.

Let $\calS(\bbR^n)$ be the Schwartz space and $\calS'(\bbR^n)$ its dual (the space of tempered distributions); let $\calE'(\bbR^n)\subset\calS'(\bbR^n)$ be the space of distributions with compact support (the dual of $C^\infty(\bbR^n)$).
We use the Fourier-transform convention $\hat u(\bm\theta)=\int_{\bbR^n}\d^n x\,u(\bm x)\,\exp(\i\bm\theta\cdot\bm x)$ for $u\in\calS'(\bbR^n)$, understood in the distributional sense.
A distribution $u\in\calE'(\bbR^n)$ is \emph{real} if $\langle u,\varphi\rangle\in\bbR$ for every real-valued $\varphi\in C^\infty(\bbR^n)$, and \emph{non-negative} (a positive measure) if $\langle u,\varphi\rangle\ge0$ for every $\varphi\ge0$.
We call $u\in\calE'(\bbR^n)$ a \emph{quasiprobability distribution} if it is real-valued only in the signed sense, i.e., it need not be non-negative, and satisfies $\langle u,1\rangle=1$.
A continuous function $g:\bbR^n\to\bbC$ is \emph{positive definite} if $\sum_{j,k=1}^{m}c_j\overline{c}_k\,g(\bm t_j-\bm t_k)\ge0$ for every finite family $\{\bm t_k\}_{k=1}^m\subset\bbR^n$ and $\{c_k\}_{k=1}^m\subset\bbC$.

To establish the extended theorem, we rely on two foundational results from functional analysis and classical probability theory.
The first is a cornerstone of the theory of distributions that characterizes the Fourier--Laplace transforms of compactly supported distributions, which will guarantee the existence and uniqueness of our quasiprobability distribution.
\begin{lemma}[Paley--Wiener--Schwartz {\cite[Thm.~7.3.1]{Hormander03}}]\label{lem:pws}
An entire function $F$ on $\bbC^n$ is the Fourier--Laplace transform of some $u\in\calE'(\bbR^n)$ with $\operatorname{supp}u\subseteq\prod_j[-R_j,R_j]$ if and only if there are $C>0$, $M\ge0$ with
\eq{
|F(\bm z)|\le C(1+|\bm z|)^{M}\exp\Big(\sum_j R_j|\Im z_j|\Big),\quad\bm z\in\bbC^n.\nonumber
}
The distribution $u$ is then unique, by injectivity of the Fourier transform on $\calS'(\bbR^n)$.
\end{lemma}
The second result is the classical standard for characterizing genuine probability measures via their Fourier transforms.
This will provide the condition linking the non-negativity of quantum quasiprobability distribution to the positive definiteness of the multivariable QCF.
\begin{lemma}[Bochner--Khinchin {\cite{Bochner55,Reed80}}]\label{lem:bk}
A continuous $g:\bbR^n\to\bbC$ with $g(\bm0)=1$ is positive definite if and only if there is a Borel probability measure $\mu$ with
\eq{
g(\bm\theta)=\int_{\bbR^n}\e^{\i\bm\theta\cdot\bm x}\d\mu(\bm x).\nonumber
}
\end{lemma}

\begin{theorem}[Extended Bochner theorem]\label{thm:extended_bochner}
Let $\bm{A}=\{A_1,\dots,A_n\}\subset\calL_{\sa}(\calH)$ be a set of observables with $R_j\coloneqq\|A_j\|<\infty$, let $N\in\bbN$ be fixed, and let $w$ satisfy $\sum_\sigma|w(\sigma)|=1$.
Then the multivariable QCF $\bm\theta\mapsto\scrC_{\bm{A}}(\bm{\theta},\rho)$ defined in Eq.~\eqref{eq:mqcf} is bounded and $C^\infty$ on $\bbR^n$, and there is a unique quasiprobability distribution $\Pr_{\bm{A}}(\,\cdot\,,\rho)\in\calE'(\bbR^n)$, with support in $\prod_j[-R_j,R_j]$, whose Fourier transform it is
\eq{\scrC_{\bm{A}}(\bm{\theta},\rho)=\int_{\bbR^n}\d^n x\,\Pr_{\bm{A}}(\bm{x},\rho)\,\e^{\i\bm{\theta}\cdot\bm{x}}.\label{eq:bochner_FT}}
This distribution has the following properties.
\begin{enumerate}
\item \emph{(Normalization)} $\int\d^n x\,\Pr_{\bm{A}}(\bm{x},\rho)=\langle\Pr_{\bm{A}}(\cdot,\rho),1\rangle=\scrC_{\bm{A}}(\bm0,\rho)=1$.
\item \emph{(Moment generation)} For every multi-index $\alpha=(\alpha_1,\dots,\alpha_n)\in\bbN_0^n$,
\eq{\int\d^n x\,\bm{x}^{\alpha}\Pr_{\bm{A}}(\bm{x},\rho)=\frac{1}{\i^{|\alpha|}}\frac{\partial^{|\alpha|}}{\partial\theta_1^{\alpha_1}\cdots\partial\theta_n^{\alpha_n}}\scrC_{\bm{A}}(\bm{\theta},\rho)\Big|_{\bm{\theta}=\bm0}.\nonumber}
\item \emph{(Reality)} $\Pr_{\bm{A}}(\,\cdot\,,\rho)$ is a real (signed) distribution if and only if
\eq{\scrC_{\bm{A}}(-\bm{\theta},\rho)=\overline{\scrC_{\bm{A}}(\bm{\theta},\rho)}\quad\text{for all }\bm\theta\in\bbR^n.\nonumber}
\item \emph{(Non-negativity)} $\Pr_{\bm{A}}(\,\cdot\,,\rho)$ is a non-negative (hence genuine) probability measure if and only if $\scrC_{\bm{A}}(\,\cdot\,,\rho)$ is a positive definite function on $\bbR^n$.
\end{enumerate}
\end{theorem}

Theorem~\ref{thm:extended_bochner} is the quasiprobability extension of Bochner's theorem.
Classical Bochner's theorem (Lemma~\ref{lem:bk}) states that a normalized continuous function is the Fourier transform of a probability measure exactly when it is positive definite.
Here the multivariable QCF need not be positive definite, and the object it transforms is a compactly supported distribution rather than a measure; existence and uniqueness come from Paley--Wiener--Schwartz (Lemma~\ref{lem:pws}), enabled by the boundedness of $\bm{A}$.
The content of the theorem is the dichotomy of statements~3 and~4: \emph{reality} of the quasiprobability is governed by the Hermitian symmetry of the multivariable QCF, whereas \emph{non-negativity}, i.e.\ the existence of a genuine probability description, is governed by its positive definiteness.
These two conditions are logically independent, and Statement~4 is precisely the point at which classical probability theory passes to quantum quasiprobability theory.
The proof is given in Appendix~\ref{sec:extended_bochner}.

We now read off the consequences for the three orderings.

\emph{Reality.}
For the symmetric orderings, the generalized unitary operator-ordering function obeys the identity $\tilde f_{\bm A}^{(N,w)}(-\bm\theta)=\tilde f_{\bm A}^{(N,w)}(\bm\theta)^\dagger$.
Indeed, for the MH ordering ($n=2$),
\eq{
\tilde f^{\text{MH}}_{A,B}(-\bm\theta)&=\tfrac12\big(\e^{-\i\theta_1 A}\e^{-\i\theta_2 B}+\e^{-\i\theta_2 B}\e^{-\i\theta_1 A}\big)\nonumber\\
&=\tilde f^{\text{MH}}_{A,B}(\bm\theta)^\dagger
}
and for the Wigner ordering
\eq{
\tilde f^{\text{W}}_{\bm A}(-\bm\theta)&=\e^{-\i\sum_j\theta_jA_j}\nonumber\\
&=\tilde f^{\text{W}}_{\bm A}(\bm\theta)^\dagger.
}
Both identities hold as operator identities, irrespective of commutativity, so $\scrC_{\bm A}(-\bm\theta,\rho)=\overline{\scrC_{\bm A}(\bm\theta,\rho)}$ for \emph{every} state.
By Statement~3, the MH- and Wigner-type quasiprobability distributions are therefore real signed distributions for every $\rho$ and every $\bm{A}$.
For the asymmetric KD ordering the operator identity fails:
\eq{
\tilde f^{\text{KD}}_{\bm A}(-\bm\theta)=\prod_{j=1}^{n}\e^{-\i\theta_jA_j},
}
while
\eq{
\tilde f^{\text{KD}}_{\bm A}(\bm\theta)^\dagger=\prod_{j=0}^{n-1}\e^{-\i\theta_{n-j}A_{n-j}},
}
and these coincide only when the factors commute.
Consequently the Hermitian symmetry of $\scrC^{\text{KD}}_{\bm A}$ is no longer automatic; whether it holds is a state-dependent condition (the reality of the KD distribution), a sufficient condition for which is $[A_j,A_k]=0$.
There is no contradiction.
That is, reality is always governed by the Hermitian symmetry of the multivariable QCF (Statement~3); this symmetry is forced by the chosen ordering for the symmetric cases, whereas for the KD ordering it must be checked for the given state and observables, and it generically fails under noncommutativity, the imaginary part being controlled by the commutators.

\emph{Non-negativity.}
If all $A_j$ commute, they share a joint spectral measure $E(\d^n x)$, the ordering is immaterial, and $\Pr_{\bm A}(\bm x,\rho)=\Tr[E(\d^n x)\rho]/\d^n x\ge0$ is the joint probability distribution; by Statement~4 the multivariable QCF is then positive definite for every state.
Thus \emph{global} commutativity of $\bm{A}$ is sufficient for classicality.
The converse is false: there exist noncommuting $\bm{A}$ and states for which $\Pr_{\bm A}$ is non-negative, and the geometry of these KD-positive states has been characterized in detail \cite{Arvidsson-Shukur21,Langrenez24,Arvidsson-Shukur24}.
Noncommutativity is therefore necessary but not sufficient for nonclassicality.

The sharp, state-dependent criterion is obtained by combining Theorem~\ref{thm:extended_bochner} with the notion of \emph{state-dependent} commutativity \cite{Ozawa05,Ozawa06,Ozawa14}.
Two observables $A,B$ \emph{commute in the state} $\rho=\ketbra{\psi}{\psi}$ if $P_A(a)P_B(b)\ket{\psi}=P_B(b)P_A(a)\ket{\psi}$ for all $a,b$; equivalently, $\ket{\psi}$ is a superposition of common eigenstates of $A$ and $B$.
A genuine joint probability distribution $\mu(a,b)$ reproducing all polynomial moments $\bra{\psi}p(A,B)\ket{\psi}$ exists if and only if $A$ and $B$ commute in $\rho$, in which case $\mu(a,b)=\bra{\psi}P_A(a)P_B(b)\ket{\psi}$.
This yields the following.
\begin{corollary}[State-dependent classicality]\label{cor:sdc}
Let $A,B\in\calL_{\sa}(\calH)$ be observables and let $\rho$ be a state.
\begin{enumerate}
\item If $A$ and $B$ commute in $\rho$, then for every ordering the quasiprobability $\Pr_{A,B}(\,\cdot\,,\rho)$ equals the joint probability distribution and is non-negative; the multivariable QCF is positive definite.
\item For the KD ordering, the non-negativity of $\Pr^{\text{KD}}_{A,B}(\,\cdot\,,\rho)$ together with reproduction of \emph{all} polynomial moments holds if and only if $A$ and $B$ commute in $\rho$.
\end{enumerate}
\end{corollary}
Thus state-dependent commutativity is strictly stronger than the non-negativity of a single ordering: commutativity in $\rho$ implies a classical description, while non-negativity of one ordering only guarantees a classical description of the moments selected by that ordering.

By a ``classical description'' we mean a non-negative, normalized distribution on the joint spectrum reproducing the relevant moments; by a ``genuine probability distribution'' we mean one satisfying the Kolmogorov axioms (non-negativity and normalization) and reproducing \emph{all} polynomial moments of $A$ and $B$.
We illustrate this asymmetry between necessary and sufficient conditions from Corollary~\ref{cor:sdc} using a simple qubit example.

\begin{example}[Qubit]\label{ex:qubit}
Let $\calH=\bbC^2$, $A=\sigma_z$, $B=\sigma_x$, which do not commute, and consider the KD ordering.

\emph{(i) Reality and non-negativity satisfied.}
For $\rho=\ketbra{0}{0}$, the KD quasiprobability takes the values
\eq{
\Pr^{\text{KD}}_{\sigma_z,\sigma_x}(+1,+1)=\Pr^{\text{KD}}_{\sigma_z,\sigma_x}(+1,-1)=\tfrac12,\nonumber
}
and zero otherwise; it is real and non-negative, so statements~3 and~4 of Theorem~\ref{thm:extended_bochner} both hold.
Nevertheless, $\ket{0}$ is not a superposition of common eigenstates of $\sigma_z$ and $\sigma_x$, i.e., $A$ and $B$ do not commute in $\rho$.
Consistently, this distribution reproduces only the KD-ordered moments and fails to give all polynomial moments, e.g. $\bra{0}\sigma_x\sigma_z\sigma_x\ket{0}=-1$, while its prediction is $\langle\sigma_z\rangle=+1$.
This explicitly demonstrates the directional logic of Corollary~\ref{cor:sdc}: while state-dependent commutativity guarantees a non-negative classical description, the mere non-negativity of a quasiprobability distribution for a specific ordering does not entail joint measurability.

\emph{(ii) Reality violated.}
For $\rho=\ketbra{+\i}{+\i}$ with $\ket{+\i}=(\ket{0}+\i\ket{1})/\sqrt2$, the KD quasiprobability is
\eq{
\Pr^{\text{KD}}_{\sigma_z,\sigma_x}(\pm1,\pm1)=\tfrac{1\mp\i}{4},\quad\Pr^{\text{KD}}_{\sigma_z,\sigma_x}(\pm1,\mp1)=\tfrac{1\pm\i}{4},\nonumber
}
which is complex.
Statement~3 of Theorem~\ref{thm:extended_bochner} fails, so the distribution is genuinely nonclassical．
\end{example}

The remaining multivariable functions, the cumulant-generating and second characteristic functions, are defined as the logarithms of $\scrM_{\bm A}$ and $\scrC_{\bm A}$, respectively.

\subsection{\label{subsec:cond-qsf}Conditional statistical functions for quantum systems}
The framework extends to systems undergoing pre- and post-selection.
We consider a system prepared in an initial state $\rho$ (pre-selection) and subsequently found to yield a specific outcome $m$ from a measurement described by a quantum instrument (post-selection).
A quantum instrument $\{\mathcal{I}_{m}\}$ acts as $\mathcal{I}_{m} (\rho) = \sum_{j} M_{mj} \rho M^{\dagger}_{mj}$, where $\{M_{mj}\}$ is a family of measurement operators \cite{Ozawa84,Ozawa04,Ozawa23}.
It satisfies $\Tr[\calI^{*}_{m}(\openone)\rho]=\Tr(\Pi_{m}\rho)$ for all $\rho$, where the family $\{\Pi_{m}\}$ is the positive operator-valued measure (POVM) defined by $\Pi_{m}\coloneqq \sum_{j}M^{\dagger}_{mj}M_{mj}$, with $\Pi_{m} \ge 0$ for all $m$ and $\sum_{m}\Pi_{m} = \openone$; it determines the probability of obtaining the outcome $m$ in the input state.
\begin{definition}[Conditional quantum moment-generating function]
For a system pre-selected in a state $\rho$ with canonical purification $\ket{\Psi}$ and post-selected by a POVM element $\Pi_{m}$, the conditional QMGF for an observable $A\in\calL_{\sa}(\calH)$ is defined as the normalized expectation value
\eq{
\scrM_{A}(\theta|\Pi_{m},\rho)\coloneqq\frac{\bra{\Psi}(\Pi_{m}\e^{\theta A} \otimes \openone_{\calH^{*}})\ket{\Psi}}{\bra{\Psi}(\Pi_{m} \otimes \openone_{\calH^{*}})\ket{\Psi}}.\label{eq:cqmfg}
}
\end{definition}
This is the QMGF for the statistics of $A$ conditioned on successful pre- and post-selection.
Unlike the unconditional single-variable case, the conditional QMGF $\scrM_{A}(\theta|\Pi_{m},\rho)$ generally takes complex values even for real $\theta$, because $A$ and $\Pi_m$ need not commute.
This is the origin of the complex character of the weak value obtained from its first derivative.
The conditional QMGF is real for all $\theta$ if and only if $\Tr(\Pi_m A^n\rho)$ is real for all $n$, and a simple sufficient condition is that \emph{any one} of the three pairs commutes:
\eq{[A,\Pi_m]=0,\quad\text{or}\quad[A,\rho]=0,\quad\text{or}\quad[\rho,\Pi_m]=0.\label{eq:cond-real}}
Indeed, if $[A,\Pi_m]=0$ then $\Pi_m\exp(\theta A)$ is self-adjoint and $\Tr(\Pi_m\exp(\theta A)\rho)$ is real; if $[A,\rho]=0$ then $\exp(\theta A)\rho$ is self-adjoint and $\Tr(\Pi_m\,\exp(\theta A)\rho)$ is real; if $[\rho,\Pi_m]=0$ then $\rho\Pi_m$ is self-adjoint and $\Tr(\exp(\theta A)\rho\Pi_m)$ is real (each uses that the trace of a product of two self-adjoint operators is real).
In the commuting case $[A,\Pi_m]=0$ the conditional QMGF reduces to the classical MGF of the conditional distribution $\Pr\{A=a|m\}=\Tr[\Pi_m P_A(a)\rho]/\Tr(\Pi_m\rho)$, and the weak value derived below coincides with the ordinary conditional expectation: this is the precise sense in which the weak value is the quantum analogue of a classical conditional expectation, justifying within our framework the interpretation advanced in Refs.~\cite{Steinberg95PRA,Steinberg95PRL,Hofmann11,Dressel14,Spriet25}.

Let the system be pre-selected in $\psi=\ketbra{\psi}{\psi}$ and post-selected onto $\phi=\ketbra{\phi}{\phi}$, corresponding to a projective measurement with $\Pi_\phi=\ketbra{\phi}{\phi}$.
The conditional QMGF is
\eq{
\scrM_{A}(\theta|\Pi_{\phi},\psi)=\frac{\bra{\phi}\e^{\theta A}\ket{\psi}}{\braket{\phi}{\psi}},
}
which generates the conditional moments of $A$ for this pre- and post-selection scheme.
Replacing the real parameter $\theta$ by the imaginary argument $-\i\theta$ turns the conditional QMGF into the conditional QCF,
\eq{
\mathcal{C}_{A}(-\theta|\Pi_{\phi},\psi)=\frac{\bra{\phi}\e^{-\i\theta A} \ket{\psi}}{\braket{\phi}{\psi}},
}
which is the modular value \cite{Kedem10}.
The framework thus contains the modular value as a special case.
The construction generalizes to several variables.
\begin{definition}[Conditional multivariable quantum moment-generating function]
For a system pre-selected in a state $\rho$ with canonical purification $\ket{\Psi}$ and post-selected by a POVM element $\Pi_{m}$, the conditional multivariable QMGF for observables $\bm{A}\subset\calL_{\sa}(\calH)$ is defined by
\eq{
\scrM_{\bm{A}}(\bm{\theta}|\Pi_{m},\rho)\coloneqq\frac{\bra{\Psi}(\Pi_{m}f_{\bm{A}}(\bm{\theta})\otimes\openone_{\calH^{*}})\ket{\Psi}}{\bra{\Psi}(\Pi_{m}\otimes\openone_{\calH^{*}})\ket{\Psi}}.\label{eq:multi-cqmfg}
}
\end{definition}
The remaining conditional quantum statistical functions are defined in the same manner.

\subsection{\label{subsec:prop}Properties of the framework}
We now derive key statistical quantities from these definitions.
Detailed proofs are given in Appendix~\ref{sec:dqsq}.
\subsubsection{\label{subsubsec:deriv_sm}Derivation of standard moments}
The first derivative of the QMGF~\eqref{eq:qmgf} at $\theta=0$ gives the standard quantum expectation value,
\eq{
\left.\frac{\d}{\d\theta}\scrM_{A}(\theta,\rho)\right|_{\theta=0}&=\Tr(A\rho)\nonumber\\
&\eqcolon\Ex_{\rho}(A).\label{eq:1st-moment}
}
For the variance, we use the centered observable $A_{0}\coloneqq A-\Ex_{\rho}(A)$.
The second derivative of its QMGF gives the variance,
\eq{
\left.\frac{\d^{2}}{\d\theta^{2}}\scrM_{A_{0}}(\theta,\rho)\right|_{\theta=0}&=\Tr\{[A-\Ex_{\rho}(A)]^{2}\rho\}\nonumber\\
&\eqcolon\Var_{\rho}(A).\label{eq:2nd-moment}
}
The framework also produces the covariance.
Using the MH-type multivariable QMGF for the centered observables $A_{0}$ and $B_{0}$, the mixed partial derivative yields the symmetrized covariance,
\eq{
&\left.\frac{\partial^{2}}{\partial\theta_{1}\partial\theta_{2}}\scrM^{\text{MH}}_{A_{0},B_{0}}(\theta_{1},\theta_{2},\rho)\right|_{\theta_{1},\theta_{2}=0}\nonumber\\
&\qquad=\frac{1}{2}\Tr[(AB+BA)\rho]-\Tr(A\rho)\Tr(B\rho)\nonumber\\
&\qquad\eqcolon\Cov_{\rho}(A,B).\label{eq:1st-moment_2params}
}
\subsubsection{\label{subsubsec:deriv_cm}Derivation of conditional moments}
The first derivative of the conditional QMGF~\eqref{eq:cqmfg} yields the weak value \cite{Aharonov88},
\eq{
\left.\frac{\d}{\d\theta}\scrM_{A}(\theta|\Pi_{m},\rho)\right|_{\theta=0}&=\frac{\Tr(\Pi_{m}A\rho)}{\Tr(\Pi_{m}\rho)}\nonumber\\
&\eqcolon\Ex_{\rho}(A|\Pi_{m}).\label{eq:1st-cond-moment}
}
The second conditional moment gives the weak variance \cite{Dressel15,Ogawa21,Matsushita24}.
For the conditionally centered observable $A'_{0}\coloneqq A-\Ex_{\rho}(A|\Pi_{m})$,
\eq{
\left.\frac{\d^{2}}{\d\theta^{2}}\scrM_{A'_{0}}(\theta|\Pi_{m},\rho)\right|_{\theta=0}&=\Ex_{\rho}(A^{2}|\Pi_{m})-\Ex_{\rho}(A|\Pi_{m})^{2}\nonumber\\
&\eqcolon\Var_{\rho}(A|\Pi_{m}).\label{eq:2nd-cond-moment}
}

\subsubsection{\label{subsubsec:gc}Structure of general correlation functions}
We here examine higher-order multivariable moments.
Our framework shows that any $n$-point correlation function can be decomposed into a chain of weak values.
The algebraic details are in Appendix~\ref{sec:dqsq}.

The $n$-th mixed partial derivative of the KD-type QMGF gives
\eq{
\left.\frac{\partial^{n}}{\partial\theta_{1}\partial\theta_{2}\cdots\partial\theta_{n}}\scrM^{\text{KD}}_{\bm{A}}(\bm{\theta},\rho)\right|_{\theta_{1},\theta_{2},\ldots,\theta_{n}=0}=\Tr(A_{1}\cdots A_{n}\rho).\label{eq:n-point_cor-func1}
}
Writing $\rho=\sum_i\lambda_i\alpha_i$ with $\alpha_{i}\coloneqq\ketbra{\alpha_{i}}{\alpha_{i}}$, and assuming nondegenerate spectra so that $P_{A_j}(a_j)=\ketbra{a_j}{a_j}$, this regroups into a chain of probabilities and weak values,
\begin{widetext}
\eq{
\left.\frac{\partial^{n}}{\partial\theta_{1}\partial\theta_{2}\cdots\partial\theta_{n}}\scrM^{\text{KD}}_{\bm{A}}(\bm{\theta},\rho)\right|_{\theta_{1},\theta_{2},\ldots,\theta_{n}=0}
&=\sum_{a_{1},\ldots,a_{n},i}a_{1}\cdots a_{n}\,\lambda_{i}\,\Pr\{A_{1}=a_{1}\Vert\alpha_{i}\}\,\prod_{k=1}^{n-1}\Ex_{\alpha_{i}}[P_{A_{k+1}}(a_{k+1})|P_{A_{k}}(a_{k})],\label{eq:n-point_cor-func2}
}
\end{widetext}
where $\Pr\{A_{1}=a_{1}\Vert\alpha_{i}\}=|\braket{a_1}{\alpha_i}|^2$ and the sequential weak value is
\eq{
\Ex_{\alpha_{i}}[P_{A_{k+1}}(a_{k+1})|P_{A_{k}}(a_{k})]\coloneqq\frac{\bra{a_{k}}P_{A_{k+1}}(a_{k+1})\ket{\alpha_{i}}}{\braket{a_{k}}{\alpha_{i}}}\label{eq:seq-weak}
}
\cite{Avdoshkin23,Guo24}.
This decomposes an $n$-point correlation function into elementary weak values, each of which is the first derivative of a conditional QMGF~\eqref{eq:cqmfg}.
The factors are not nested in a way that requires a single sequential weak measurement: each weak value $\Ex_{\alpha_i}[P_{A_{k+1}}(a_{k+1})|P_{A_k}(a_k)]$ refers to its own pre- and post-selection and can be obtained from an independently prepared ensemble, so the decomposition supports both sequential and parallel (independently realized) weak-measurement protocols for assembling correlation functions \cite{Dressel18,Del-Re24,Emori25}.
By the cyclic property of the trace, the starting point of the decomposition is arbitrary, giving distinct but equivalent schemes.

\section{\label{sec:geometry}Geometry and convexity of the operator-ordering parameter}
The parameter $N$ in the generalized operator-ordering function $f_{\bm{A}}^{(N, w)}(\bm{\theta})$ is not only a tool for Trotterization; it indexes how noncommutativity is treated as one passes from the product form ($N=1$) to the exponential-sum form ($N\to\infty$).
We collect the structural properties that we can establish rigorously.

\subsection{\label{subsec:convexity}Convexity and inequalities}

Because the single-variable QMGF is mathematically equivalent to the classical MGF of a genuine probability distribution [Eq.~\eqref{eq:qmgf_classical}], it inherits all standard classical convexity properties.
For any observable $A\in\calL_{\sa}(\calH)$ and state $\rho$, the QMGF $\scrM_{A}(\theta,\rho)$ is strictly positive, convex, and log-convex as a function of $\theta \in \bbR$.
For instance, applying H{"{o}}lder's inequality and the positivity of convex combinations of exponential functions yields the log-convexity condition: $\scrM_{A}(\lambda\theta_1+(1-\lambda)\theta_2,\rho)\le\scrM_{A}(\theta_1,\rho)^{\lambda}\scrM_{A}(\theta_2,\rho)^{1-\lambda}$ for $0\le\lambda\le1$.
Furthermore, the Cauchy--Schwarz and Jensen's inequalities provide the lower bounds $\scrM_{A}(\theta,\rho)\,\scrM_{A}(-\theta,\rho)\ge1$ and $\scrM_{A}(\theta,\rho)\ge\exp[\theta\Ex_\rho(A)]$, respectively.
The QCGF $\scrK_{A}(\theta,\rho)$ is consequently convex, with its second derivative $\scrK_{A}''(\theta,\rho)=\Var_{\Pr_\theta}(A)\ge0$ representing the variance under the exponentially tilted distribution $\Pr_\theta(a)\propto\exp(\theta a)\Pr\{A=a\Vert\rho\}$.

Beyond the single-variable case, a genuinely multivariable convexity statement can be established at the symmetric limit ($N\to\infty$) for the maximally mixed state (the tracial case).
For a set of observables $\bm{A}\subset\calL_{\sa}(\calH)$ on a $d$-dimensional space and the maximally mixed state $\rho=\openone/d$, the Wigner-type multivariable QCGF
\eq{
\scrK^{\text{W}}_{\bm{A}}(\bm{\theta})=\log\Big[\tfrac{1}{d}\Tr\,\e^{\sum_j\theta_jA_j}\Big]\nonumber
}
is a convex function on $\bbR^n$.
This is fundamentally a manifestation of the convexity of the log-partition function $\bm\theta\mapsto\log\Tr\,\exp[H(\bm\theta)]$ for the affine family $H(\bm\theta)=\sum_j\theta_jA_j$.
This property is a standard consequence of the Peierls--Bogoliubov inequality, or equivalently, H{"{o}}lder's inequality applied to Schatten norms in conjunction with the Golden--Thompson inequality (see, e.g., Refs.~\cite{Lieb73,Carlen10}).

\subsection{\label{subsubsec:gt}Golden--Thompson comparison}
The transition from the product form ($N=1$) to the exponential-sum form ($N\to\infty$) is constrained, at infinite temperature, by the Golden--Thompson inequality \cite{Golden65,Thompson65}
\eq{
\Tr(\e^{H+K})\le\Tr(\e^{H}\e^{K})\label{eq:gt}
}
for observables $H,K\in\calL_{\sa}(\calH)$.
With a state $\rho=\openone/d$, one has $\scrM^{\text{MH}}_{\bm A}(\bm\theta,\rho)=\Re\Tr[\exp(\theta_1A_1)\exp(\theta_2A_2)]/d=\Tr[\exp(\theta_1A_1)\exp(\theta_2A_2)]/d$ and $\scrM^{\text{W}}_{\bm A}(\bm\theta,\rho)=\Tr[\exp(\theta_1A_1+\theta_2A_2)]/d$, so Eq.~\eqref{eq:gt} gives
\eq{
\scrM^{\text{W}}_{\bm{A}}(\bm{\theta},\rho)\le\scrM^{\text{MH}}_{\bm{A}}(\bm{\theta},\rho).\label{eq:gt-bound}
}
For a general state, Eq.~\eqref{eq:gt-bound} does \emph{not} hold; a Taylor expansion shows that $\scrM^{\text{W}}_{\bm A}$ and $\scrM^{\text{MH}}_{\bm A}$ agree through second order in $\bm\theta$ (so they reproduce the same mean and covariance), and their difference, which first appears at third order, has no definite sign in general.

The relevant tools for partial general-state statements are the multivariate Golden--Thompson extensions of the Araki--Lieb--Thirring inequality \cite{Lieb73,Ando94} and the multivariate trace inequalities of Ref.~\cite{Sutter17}, but these do not yield Eq.~\eqref{eq:gt-bound} for arbitrary $\rho$.
The two endpoints have a clear reading: the Wigner limit fuses the noncommuting generators into a single exponential before exponentiation, whereas the KD limit preserves the group structure of the individual exponentials, reflecting a measurement sequence in which noncommutativity is temporally or spatially resolved.
A property relevant to information measures is the monotonicity of the operator geometric mean under positive linear maps $\Phi$, $\Phi(A\#B)\le\Phi(A)\#\Phi(B)$, which follows from the Kubo--Ando theory \cite{Kubo80}, with the related log-majorization due to Ref.~\cite{Ando94}.

\section{\label{sec:app}Application: Quantum parameter estimation}
Having established the properties of the quantum statistical functions and their relation to quasiprobability distributions, we turn to an application.
We show how the framework supports quantum parameter estimation by importing the classical method of moments (MM) and generalized method of moments (GMM) \cite{Hansen82} into quantum metrology.
We introduce the quantum method of moments (QMM) and the quantum generalized method of moments (QGMM) as alternatives to full quantum state tomography and to maximum likelihood estimation \cite{Hradil97}.
These methods use the QMGF $\scrM_{\bm{A}}(\bm{\theta},\rho)$ to estimate parameters encoded in a quantum state.

\subsection{\label{subsec:qmm}Quantum method of moments}
Consider a quantum system described by a parameterized density operator $\rho_{\bm{\phi}}$, where $\bm{\phi} = (\phi_1, \dots, \phi_K) \in \Phi \subseteq \mathbb{R}^K$ is a vector of unknown physical parameters (e.g., Hamiltonian coefficients, temperature, or diffusion constants).
The goal is to estimate $\bm{\phi}$ from measurements.
In the classical method of moments, one equates the theoretical moments of a distribution to the empirical moments obtained from data.
While the classical framework often uses higher-order moments of a single random variable (e.g., $\Ex[X], \Ex[X^2], \dots$), the quantum setting allows one to choose a set of distinct observables (e.g., noncommuting Pauli operators $X, Y, Z$) or powers of a single observable; the framework accommodates both.
Let $\calO = \{O_1, \dots, O_K\}$ be a chosen set of observables (moment conditions) designed to identify the parameters.
The theoretical expectations $\mu_k(\bm{\phi})$ are generated by the first derivatives of the QMGF [Eq.~\eqref{eq:qmgf}] or the multivariable QMGF [Eq.~\eqref{eq:mqmgf}] for the hypothesis state $\rho_{\bm{\phi}}$,
\eq{
\mu_k(\bm{\phi}) &\coloneqq \Ex_{\rho_{\bm{\phi}}}(O_k).
}
If $O_k$ corresponds to a higher moment, it is obtained from the $n$-th derivative of $\scrM_{A}(\theta, \rho_{\bm{\phi}})$.
Suppose we perform $M$ independent measurements, yielding empirical averages $\hat{\mu}_k$ for each observable $O_k$.
The QMM estimator $\hat{\bm{\phi}}_{\text{QMM}}$ is the solution of the system of $K$ equations
\eq{
\bm{\mu}(\hat{\bm{\phi}}_{\text{QMM}}) = \hat{\bm{\mu}}, \label{eq:qmm_system}
}
where $\bm{\mu} = (\mu_1, \dots, \mu_K)^{T}$ and $\hat{\bm{\mu}} = (\hat{\mu}_1, \dots, \hat{\mu}_K)^{T}$.
The QMGF ensures that $\bm{\mu}(\bm{\phi})$ is well-defined and smooth, so root-finding algorithms (e.g., Newton--Raphson) can solve Eq.~\eqref{eq:qmm_system}.

\subsection{\label{subsec:qgmm}Quantum generalized method of moments}
The QMM is suited to minimal setups but is limited when one measures more observables than necessary.
In quantum Hamiltonian learning \cite{Wiebe14,Wang17} and state tomography \cite{James01,Paris04}, the system is often over-identified: the number of accessible moment conditions $L$ exceeds the number of unknown parameters $K$, i.e., $L > K$.
In this regime, the system~\eqref{eq:qmm_system} generally has no solution.
The QGMM estimates parameters by minimizing the distance between the theoretical and empirical moments in a metric set by the quantum statistics of the system.
Let $\bm{g}(\bm{\phi}) \coloneqq \hat{\bm{\mu}} - \bm{\mu}(\bm{\phi})$ be the vector of moment conditions.
The QGMM estimator $\hat{\bm{\phi}}_{\text{QGMM}}$ minimizes the quadratic objective
\eq{
\hat{\bm{\phi}}_{\text{QGMM}} &= \operatorname*{argmin}_{\bm{\phi} \in \Phi} J(\bm{\phi}), \\
J(\bm{\phi}) &= \bm{g}(\bm{\phi})^{T} \mathbf{W} \bm{g}(\bm{\phi}), \label{eq:qgmm_obj}
}
where $\mathbf{W}$ is an $L \times L$ positive definite weighting matrix.
The choice of $\mathbf{W}$ controls the asymptotic efficiency of the estimator.
In classical GMM, the optimal weighting matrix is the inverse of the covariance matrix of the moment conditions \cite{Hansen82}, $\mathbf{W}_{\text{opt}}=\mathbf{\Sigma}^{-1}$, and the resulting estimator attains the minimal asymptotic covariance $(D^T\mathbf{\Sigma}^{-1}D)^{-1}$, where $D=\nabla_{\bm\phi}\bm\mu$ (see Appendix~\ref{sec:dqsq}).
The proposed framework lets us compute the matrix $\mathbf{\Sigma}$ for the quantum system.
Using the MH-type ordering of Eq.~\eqref{eq:mh_dist}, we construct the quantum covariance matrix $\mathbf{\Sigma}(\bm{\phi})$ between the observables $O_j$ and $O_k$:
\eq{
\Sigma_{jk}(\bm{\phi}) &\coloneqq \Cov_{\rho_{\bm{\phi}}}(O_j, O_k)\nonumber\\
&=\frac{\partial^2}{\partial\theta_j\partial\theta_k}\scrM^{\text{MH}}_{\bm{O}}(\bm\theta,\rho_{\bm\phi})\Big|_{\bm\theta=\bm0},\label{eq:Sigma_def}
}
that is, $\Sigma_{jk}=\tfrac12\Tr(\{O_j,O_k\}\rho_{\bm\phi})-\mu_j\mu_k$ obtained from the mixed derivatives of the MH-type QMGF.

The QGMM estimator is constructed as follows.
\begin{enumerate}
    \item Obtain an initial consistent estimator $\bar{\bm{\phi}}$ (e.g.\ using the QMM estimator using a just-identified subset, equivalently $\mathbf{W}=\openone$).
    \item Compute the optimal weighting matrix $\mathbf{W}_{\text{opt}} = [\mathbf{\Sigma}(\bar{\bm{\phi}})]^{-1}$ using the derivatives of the multivariable QMGF evaluated at $\bar{\bm{\phi}}$.
    \item Minimize Eq.~\eqref{eq:qgmm_obj} with $\mathbf{W}_{\text{opt}}$ to obtain the final estimate.
\end{enumerate}
The QGMM is useful when the full likelihood is computationally intractable (e.g., requiring full state diagonalization) but the QMGF is accessible through approximations.

\subsection{\label{subsec:example}Example: transverse-field Ising model}
To illustrate QMM and QGMM, we consider the one-dimensional transverse-field Ising model on a ring of $N$ spins, with Hamiltonian
\eq{
H = -\sum_{i} J_{i} \sigma_{i}^{z}\sigma_{i+1}^{z} - \sum_{i} h_{i} \sigma_{i}^{x}.
}
Assuming translation invariance, the parameters reduce to a uniform interaction strength $J$ and a uniform transverse field $h$, so $\bm{\phi} = (J, h)^{T}$.
The system is in thermal equilibrium
\eq{
\rho_{\bm{\phi}} = \frac{\e^{-\beta H(\bm{\phi})}}{Z},
}
at a known inverse temperature $\beta$.
For explicit solutions, we work in the high-temperature regime ($\beta \ll 1$), where we expand $\rho_{\bm\phi}=2^{-N}(\openone-\beta H+\tfrac{\beta^2}{2}H^2-\cdots)$ and retain the order needed for each quantity.

\subsubsection{Estimation via QMM}
With $K=2$ unknown parameters, we use two moment conditions, the nearest-neighbor correlation and the transverse magnetization,
\eq{
O_1 &= \frac{1}{N} \sum_{i=1}^N \sigma_{i}^{z}\sigma_{i+1}^{z}, \\
O_2 &= \frac{1}{N} \sum_{i=1}^N \sigma_{i}^{x}.
}
In the high-temperature approximation, the theoretical moments $\mu_k(\bm{\phi}) = \Tr(O_k \rho_{\bm{\phi}})$ are
\eq{
\mu_1(J, h) &\approx \frac{1}{2^N} \Tr\left[ O_1 \Big(\openone + \beta J \sum_j \sigma_j^z \sigma_{j+1}^z + \beta h \sum_j \sigma_j^x\Big) \right] \nonumber\\
&= \beta J, \\
\mu_2(J, h) &\approx \frac{1}{2^N} \Tr\left[ O_2 \Big(\openone + \beta J \sum_j \sigma_j^z \sigma_{j+1}^z + \beta h \sum_j \sigma_j^x\Big) \right] \nonumber\\
&= \beta h.
}
Given the empirical averages $\hat{\mu}_1$ and $\hat{\mu}_2$, the QMM estimators solve $\mu_k(\hat{\bm{\phi}}) = \hat{\mu}_k$,
\eq{
\begin{cases}
\beta \hat{J}_{\text{QMM}} = \hat{\mu}_1 \\
\beta \hat{h}_{\text{QMM}} = \hat{\mu}_2
\end{cases}
\implies
\hat{J}_{\text{QMM}} = \frac{\hat{\mu}_1}{\beta}, \quad \hat{h}_{\text{QMM}} = \frac{\hat{\mu}_2}{\beta},
}
a direct inversion of the moment functions for the Hamiltonian parameters.

\subsubsection{Estimation via QGMM}
Suppose we measure an additional observable, the next-nearest-neighbor correlation
\eq{
O_3 = N^{-1} \sum_{i} \sigma_{i}^{z}\sigma_{i+2}^{z},
}
so the system is over-identified ($L=3 > K=2$).
At first order in the high-temperature expansion, $\Ex_{\rho_{\bm{\phi}}}(O_3)$ vanishes, so we keep the second-order term $\rho_{\bm{\phi}} \approx 2^{-N}(\openone - \beta H + \beta^{2}H^{2}/2)$.
The theoretical moment for $O_3$ arises from the cross-term in $H^2$,
\eq{
\mu_3(J, h) &\approx \frac{1}{2^N} \Tr\left[ O_3 \frac{\beta^2}{2} \Big( J \sum_j \sigma_j^z \sigma_{j+1}^z \Big)^2 \right] \nonumber\\
&= (\beta J)^2.\label{eq:mu3}
}
The vector of moment conditions is $\bm{g}(J, h) = (\hat{\mu}_1 - \beta J, \hat{\mu}_2 - \beta h, \hat{\mu}_3 - (\beta J)^2)^{T}$.
To construct the optimal estimator, we evaluate the quantum covariance matrix $\mathbf{\Sigma}$ to determine $\mathbf{W}_{\text{opt}} = \mathbf{\Sigma}^{-1}$.
The leading contribution comes from the identity component $\openone/2^{N}$ of $\rho$.
For the diagonal elements, we have
\eq{
\Sigma_{kk} &\approx \frac{1}{2^N} \Tr(O_k^2) \nonumber\\
&= \frac{1}{N},
}
where only terms with $i=j$ survive the trace.
For the off-diagonal elements, the trace $\Tr(O_i O_j)$ vanishes because the Pauli indices cannot match to form the identity, and the first nonvanishing contribution, from the $\beta H$ term in $\rho$, is of order $O(\beta/N)$, negligible compared to the variance term $O(1/N)$.
The covariance matrix is therefore proportional to the identity at leading order,
\eq{
\mathbf{\Sigma} \approx \frac{1}{N} \openone \implies \mathbf{W}_{\text{opt}} \approx N \openone.
}
The QGMM estimator minimizes $J(\bm{\phi}) = \bm{g}^{T} \mathbf{\Sigma}^{-1} \bm{g}$.

We use the one-step (linearized) GMM update, i.e.\ a single Gauss--Newton step toward $\operatorname{argmin}J(\bm\phi)$ in Eq.~\eqref{eq:qgmm_obj} with $\mathbf{W}=\mathbf\Sigma^{-1}$, starting from the QMM estimate $\bar{\bm\phi}=(\hat\mu_1/\beta,\hat\mu_2/\beta)^T$.
We start from $\bar{\bm\phi}$ rather than from a fit including $O_3$ because $(O_1,O_2)$ already exactly determine $(J,h)$ at leading order; $\hat\mu_3$ then enters only through the update, refining the estimate via the over-identifying condition.
The Jacobian at $\bar{\bm\phi}$ (using $\hat\mu_1=\beta\bar J$) is
\eq{
D&=\nabla_{\bm\phi}\bm\mu\big|_{\bar{\bm\phi}}\nonumber\\
&\approx\begin{pmatrix}\beta&0\\0&\beta\\2\beta\hat\mu_1&0\end{pmatrix},
}
since $\partial\mu_3/\partial J=2\beta^2 J$ evaluated at $\bar J=\hat\mu_1/\beta$ gives $2\beta\hat\mu_1$.
The one-step update therefore is
\eq{\hat{\bm\phi}_{\text{QGMM}}=\bar{\bm\phi}+(D^T\mathbf\Sigma^{-1}D)^{-1}D^T\mathbf\Sigma^{-1}\big(\hat{\bm\mu}-\bm\mu(\bar{\bm\phi})\big).\label{eq:onestep}}
Since the vector of theoretical moments is $\bm\mu(\bar{\bm\phi})=(\hat\mu_1,\hat\mu_2,\hat\mu_1^2)^T$, the residual is $\hat{\bm\mu}-\bm\mu(\bar{\bm\phi})=(0,0,\hat\mu_3-\hat\mu_1^2)^T$.
A short computation gives
\eq{
\hat{J}_{\text{QGMM}}&=\frac{\hat\mu_1}{\beta}+\frac{2\hat\mu_1}{\beta(1+4\hat\mu_1^2)}\Big(\hat\mu_3-\hat\mu_1^2\Big),\label{eq:Jqgmm}\\
\hat{h}_{\text{QGMM}}&=\frac{\hat\mu_2}{\beta}.\label{eq:hqgmm}
}
The update corrects the initial estimate using the consistency between the observed next-nearest-neighbor correlation $\hat\mu_3$ and its prediction $\hat\mu_1^2=(\beta\bar J)^2$ from Eq.~\eqref{eq:mu3}; $\hat h$ is unchanged because $O_3$ carries no information about $h$ at this order.

The asymptotic variance of the optimal estimator is the $(1,1)$ entry of $(D^T\mathbf\Sigma^{-1}D)^{-1}$.
With $D^T\mathbf\Sigma^{-1}D=N\,\mathrm{diag}\big(\beta^2(1+4\hat\mu_1^2),\beta^2\big)$,
\eq{
\Var(\hat{J}_{\text{QGMM}})=\frac{1}{N\beta^2(1+4\hat\mu_1^2)},
}
to be compared with $\Var(\hat{J}_{\text{QMM}})=1/(N\beta^2)$.
The ratio is $(1+4\hat\mu_1^2)^{-1}\le1$.
The over-identifying condition $O_3$ can only reduce the variance, so the QGMM is never less efficient than the QMM, and the reduction is the efficiency gain quantified by the GMM bound.
The gain is governed by the coupling strength: as $\hat\mu_1=\beta\bar J\to0$ the factor tends to $1$ and the gain vanishes (a weakly coupled $O_3$ carries little information about $J$), while for $\hat\mu_1=O(1)$ the reduction is substantial.
That this is the minimal achievable variance, not merely a reduction, is the content of the GMM efficiency bound $(D^T\mathbf\Sigma^{-1}D)^{-1}$, verified for this example in Appendix~\ref{sec:qgmm}.

\section{\label{sec:alternative}Alternative constructions}
We compare our definition with several alternatives, clarifying first which apparent alternatives are in fact equivalent to ours and which are genuinely different.

\subsection{\label{subsec:naive}Equivalent and naive exponential forms}
Several candidate generating functions coincide with ours or differ from it only trivially, by the cyclic property of the trace.
In statistical mechanics one writes generating functions as $\Tr[\exp(-\beta H)\exp(\theta A)]$; dividing by the partition function $Z=\Tr\,\exp(-\beta H)$ gives $\Tr[\exp(-\beta H)\exp(\theta A)]/Z=\Tr[\rho_{\text{Gibbs}}\exp(\theta A)]=\scrM_{A}(\theta,\rho_{\text{Gibbs}})$, our definition for the Gibbs state.
The symmetric ``sandwich'' form is likewise equal to ours, $\Tr[\exp(\theta A/2)\rho\,\exp(\theta A/2)]=\Tr[\exp(\theta A)\rho]=\scrM_{A}(\theta,\rho)$, so it reproduces the standard moments and is not a distinct construction.
The doubled exponential merely rescales the argument, $\Tr[\exp(\theta A)\rho\,\exp(\theta A)]=\Tr[\exp(2\theta A)\rho]=\scrM_{A}(2\theta,\rho)$, so its derivatives give $2\Ex_\rho(A)$ and (for the centered observable) $4\Var_\rho(A)$, deviating from the standard quantities only by constant factors.
None of these is genuinely new.
The distinct constructions, treated below, are the geometric-mean construction and the information-spectrum functional.
We also note that, setting $A=H+\theta V$ with $V$ a source and $\rho\propto\openone$, Eq.~\eqref{eq:mqmgf} extends Eq.~\eqref{eq:partition_func}, providing a more flexible setting for the generating functional $Z[J]$.

\subsection{\label{subsec:gmb}Geometric mean-based statistical functions}
Using the metric structure of the matrix geometric mean introduced in Sec.~\ref{subsec:mgm}, we define a QMGF that measures the geometric overlap between a reference state and a perturbed state.
We consider the overlap between $\rho$ and the exponential operator $\sigma_{\theta} = \exp(\theta V)$,
\eq{
\scrM^{\text{geo}}_{V}(\theta, \rho) &\coloneqq \Tr\left( \rho \# \e^{\theta V} \right) \nonumber\\
&= \Tr\left[ \rho^{1/2} \left( \rho^{-1/2} \e^{\theta V} \rho^{-1/2} \right)^{1/2} \rho^{1/2} \right].
}
In the amplitude language of Sec.~\ref{subsubsec:amp}, with the canonical amplitude $W_\rho=\sqrt\rho$ and the parallel amplitude $W_\sigma(\theta)$, this is the overlap of parallel amplitudes
\eq{
\scrM^{\text{geo}}_{V}(\theta, \rho) = \Tr\left[ W_\rho W_\sigma(\theta)^\dag \right],
}
i.e.\ the inner product of the corresponding purifications.
We derive the derivatives of $\scrM^{\text{geo}}_{V}(\theta, \rho)$ with respect to $\theta$ (see Appendix~\ref{sec:dgmbf} for more detail).
The first derivative is
\eq{
\left. \frac{\d}{\d\theta} \scrM^{\text{geo}}_{V}(\theta, \rho) \right|_{\theta=0} = \frac{1}{2}\Tr(\sqrt{\rho} V),\label{eq:1st-moment_gm}
}
the $\sqrt{\rho}$-weighted value.
The second derivative at $\theta=0$ is
\eq{
\left. \frac{\d^{2} \scrM^{\text{geo}}_{V}(\theta, \rho)}{\d \theta^{2}} \right|_{\theta=0} &= \frac{1}{2}\Tr(\sqrt{\rho} V^2)\nonumber\\
&\quad\mbox{}- \sum_{k,j} \frac{p_k^{3/2}}{(\sqrt{p_k} + \sqrt{p_j})^2} |\bra{k}V\ket{j}|^2.\label{eq:2nd-moment_gm}
}
The first derivative differs from the standard expectation value $\Tr(\rho V)$ in general, reflecting the $\sqrt\rho$ weighting intrinsic to the geometric mean.
Hence the geometric-mean construction does not reproduce the standard quantum moments.

\subsection{\label{subsec:is}Information-spectral statistical functions}
Complementary to the matrix geometric mean, the information-spectrum method focuses on the distribution of the log-likelihood \cite{Nagaoka07,Audenaert07}.
In quantum hypothesis testing between a null hypothesis state $\rho$ and an alternative hypothesis state $\sigma$, the central object is the spectral decomposition of the difference of their log-likelihoods.
The associated QMGF, the quantum Chernoff functional, is
\eq{
\psi(\theta) \coloneqq \log \Tr\left( \rho^{1-\theta} \sigma^{\theta} \right)
}
for $\theta \in [0, 1]$.
Its derivatives at $\theta=0$ give the information-theoretic quantities governing the asymptotic distinguishability.
The first derivative equals minus the quantum relative entropy $D(\rho\|\sigma)$,
\eq{
\left. \frac{\d \psi(\theta)}{\d \theta} \right|_{\theta=0} &= \Tr\left[ \rho (\log \sigma - \log \rho) \right] \nonumber\\
&\eqcolon -D(\rho\|\sigma),\label{eq:1st-moment_is}
}
the center of the information spectrum.
The second derivative defines the quantum relative entropy variance $\Var(\rho\|\sigma)$, which controls the dispersion of the hypothesis-testing error in the second-order asymptotics,
\eq{
\left. \frac{\d^{2} \psi(\theta)}{\d \theta^{2}} \right|_{\theta=0} &= \Tr\left[ \rho (\log \sigma - \log \rho)^2 \right] - \left( \Tr[\rho (\log \sigma - \log \rho)] \right)^2 \nonumber \\
&\eqcolon \Var(\rho\|\sigma),\label{eq:2nd-moment_is}
}
the width of the information spectrum.
This functional generates a different quantities from ours; obtaining the standard statistical quantities instead requires the generalized operator-ordering function of Sec.~\ref{subsubsec:multi}.

\section{\label{sec:conclusion}Conclusion}
We have proposed a unified framework for statistical functions in quantum systems.
Using the generalized operator-ordering function and its unitary variant, we introduced the quantum moment-generating, characteristic, cumulant-generating, and second characteristic functions, together with their multivariable and conditional extensions, and made the role of operator ordering explicit.
The single-variable functions reduce exactly to the classical statistical functions of the projective-measurement distribution, the multivariable functions reduce to this case when the relevant observables commute, and the conditional functions generate the weak value and weak variance, with the weak value reducing to the ordinary conditional expectation when the post-selection commutes with the observable in the relevant state.
The extended Bochner theorem identifies the failure of positive definiteness of the multivariable quantum characteristic function as the boundary between classical probability and quantum quasiprobability, and, combined with the state-dependent notion of commutativity, shows that state-dependent commutativity is strictly stronger than the non-negativity of any single ordering.
We established convexity of the single-variable quantum cumulant-generating function and of the Wigner-type quantum cumulant-generating function at infinite temperature, and a Golden--Thompson comparison between orderings.
As an application, we imported the method of moments and the generalized method of moments into quantum estimation, obtaining the QMM and QGMM estimators, worked out for the transverse-field Ising model, where the over-identifying moment condition yields a quantifiable efficiency gain.
We note that independent work on a related topic has recently been carried out by Jordan, Arvidsson-Shukur, and Steinberg \cite{Jordan26}.

\begin{acknowledgments}
The author thanks David R. M. Arvidsson-Shukur, Anna Jen{\v{c}}ov{\'{a}}, Andrew N. Jordan, Seok Hyung Lie, Takato Mori, Masanao Ozawa, Aephraim M. Steinberg, and Akihisa Tomita for helpful discussions.
This work was supported by the Japan Science and Technology Agency (JST) as part of Adopting Sustainable Partnerships for Innovative Research Ecosystem (ASPIRE) Grant No. JPMJAP2318, JST SPRING Grant No. JPMJSP2119, and the RIKEN Junior Research Associate Program.
\end{acknowledgments}

\appendix

\section{\label{sec:dwf}Quantum characteristic function and discrete Wigner functions}
In this appendix, we describe the connection between the Wigner-type multivariable QCF $\scrC_{\bm{A}}^{\text{W}}$ derived from the generalized unitary operator-ordering function~\eqref{eq:gen_uni_func} and the theory of discrete Wigner functions for finite-dimensional systems \cite{Wootters87,Gross06,DeBrota20}.
Consider a $d$-dimensional Hilbert space $\calH \cong \bbC^d$.
The discrete phase space is the grid $\calV \coloneqq \bbZ_d \times \bbZ_d$.
Let $\{ \ket{k} \}_{k \in \bbZ_d}$ be the computational basis.
We introduce the generalized Pauli operators, the shift operator $X$ and the clock operator $Z$, defined by
\eq{
X\ket{k} &= \ket{k+1 \bmod d},\\
Z\ket{k} &= \omega^k \ket{k},
}
where $\omega = \exp(2\pi \i / d)$ is a primitive root of unity.
These satisfy the Weyl relation $ZX = \omega XZ$.
For a point $\bm{u} = (q, p) \in \calV$, the displacement operator $D_{\bm{u}}$ is
\eq{
D_{\bm{u}} = \tau^{-qp} Z^{q} X^{p},
}
where $\tau$ is a phase ensuring the required covariance.
For odd $d$, one chooses $\tau = \exp[(d+1)\pi \i / d]$ (so that $\tau^2 = \omega$); for even $d$, the phase requires the extended phase space $\bbZ_{2d} \times \bbZ_{2d}$, but the algebraic structure is analogous.
The operators $D_{\bm{u}}$ form an operator basis for $\calL(\calH)$ with $\Tr(D_{\bm{u}}^\dagger D_{\bm{v}}) = d \,\delta_{\bm{u}, \bm{v}}$.
Connecting this to the generalized unitary operator-ordering function $\tilde{f}_{\bm{A}}^{(N, w)}(\bm{\theta})$, in the discrete setting the continuous parameters $\bm{\theta}$ are replaced by the discrete phase-space coordinates $\bm{u} \in \calV$, and the observables $\bm{A}$ are identified with the generators of the finite Heisenberg--Weyl group.
The Wigner-type limit ($N \to \infty$) yields the exponential of the sum of generators, i.e., the Weyl-ordered product, which in the discrete domain is the displacement operator $D_{\bm{u}}$.
Thus the Wigner-type QCF is the expectation value of the displacement operator,
\eq{
\scrC_{\bm{A}}^{\text{W}}(\bm{u}, \rho) \longleftrightarrow \Tr(\rho D_{\bm{u}}) \eqcolon \chi_{\rho}(\bm{u}),
}
the discrete QCF.
The discrete Wigner function $W_{\rho}(\bm{x})$ at $\bm{x} = (x_1, x_2) \in \calV$ is the symplectic Fourier transform of the characteristic function,
\eq{
W_{\rho}(\bm{x}) = \frac{1}{d} \sum_{\bm{u} \in \calV} \omega^{\{\bm{x}, \bm{u}\}} \chi_{\rho}(\bm{u}),
}
where $\{\bm{x}, \bm{u}\} = x_1 u_2 - x_2 u_1$ is the symplectic inner product on $\calV$.
This transform is invertible, giving a duality between the Wigner-function and characteristic-function representations.
The phase-point operators $A_{\bm{x}}$ (Fano operators in the continuous case), with $W_{\rho}(\bm{x}) = \Tr(\rho A_{\bm{x}})$, are
\eq{
A_{\bm{x}} = \frac{1}{d} \sum_{\bm{u} \in \calV} \omega^{\{\bm{x}, \bm{u}\}} D_{\bm{u}},
}
and satisfy the Stratonovich--Weyl postulates generalized to finite dimensions, including Hermiticity and covariance.
The Wigner-type limit ($N \to \infty$) of our framework therefore generates the discrete QCF $\chi_{\rho}(\bm{u})$, which lies in the dual domain of the Wigner distribution, linking the formalism to the established theory of discrete Wigner functions.

\section{\label{sec:extended_bochner}Proof of the extended Bochner theorem}
We prove Theorem~\ref{thm:extended_bochner} by direct construction, treating the four assertions in turn after establishing the analytic estimate on which the construction rests.

\begin{proofof}{Theorem~\ref{thm:extended_bochner}}
Throughout, $R_j\coloneqq\|A_j\|<\infty$, $N$ is fixed, and $\sum_\sigma|w(\sigma)|=1$.
For $u\in\calE'(\bbR^n)$ we write $\hat u(\bm\theta)=\langle u,\exp[\i\bm\theta\cdot(\cdot)]\rangle$ for its Fourier transform; the complex conjugate $\bar u$ is defined by $\langle\bar u,\varphi\rangle=\overline{\langle u,\bar\varphi\rangle}$ for $\varphi\in C^\infty$.

\emph{Analytic estimate.}
For a bounded self-adjoint $A_j$ with spectral measure $\d P_{A_j}(\lambda)$ on $\operatorname{spec}(A_j)\subset[-R_j,R_j]$ and $z_j\in\bbC$, the spectral theorem gives $\exp(\i z_j A_j)=\int_{\operatorname{spec}(A_j)}\exp(\i z_j\lambda)\,\d P_{A_j}(\lambda)$, which is entire in $z_j$.
By the continuous functional calculus for self-adjoint operators \cite[Ch.~VII]{Reed80}, $\|f(A_j)\|=\sup_{\lambda\in\operatorname{spec}(A_j)}|f(\lambda)|$, so with $f(\lambda)=\exp(\i z_j\lambda)$, $|f(\lambda)|=\exp(-\lambda\Im z_j)$ and
\eq{
\big\|\e^{\i z_j A_j}\big\|=\sup_{\lambda\in\operatorname{spec}(A_j)}\e^{-\lambda\Im z_j}\le\e^{R_j|\Im z_j|}.\label{eq:normbound}
}
Hence each factor $\exp(\i z_{\sigma(j)}A_{\sigma(j)}/N)$ in Eq.~\eqref{eq:gen_uni_func} is entire with norm at most $\exp(R_{\sigma(j)}|\Im z_{\sigma(j)}|/N)$.
For a fixed $\sigma$, by submultiplicativity of the operator norm,
\eq{
\Big\|\prod_{j=1}^{n}\e^{\i z_{\sigma(j)}A_{\sigma(j)}/N}\Big\|&\le\prod_{j=1}^{n}\e^{R_{\sigma(j)}|\Im z_{\sigma(j)}|/N}\nonumber\\
&=\prod_{k=1}^{n}\e^{R_k|\Im z_k|/N},\nonumber
}
where the last equality holds because the product runs over all $n$ scalar factors and is therefore independent of the ordering $\sigma$ (it is a reindexing of the same factors).
Summing over $S_n$ and using $\sum_\sigma|w(\sigma)|=1$, the inner bracket of $\tilde f_{\bm A}^{(N,w)}$ has norm at most $\prod_k\exp(R_k|\Im z_k|/N)$, and its $N$-th power has norm at most $\prod_k\exp(R_k|\Im z_k|)$.
Thus $\bm z\mapsto\tilde f_{\bm A}^{(N,w)}(\bm z)$ is an entire $\calL(\calH)$-valued function, the scalar function $\scrC_{\bm A}(\bm z,\rho)=\Tr[\tilde f_{\bm A}^{(N,w)}(\bm z)\rho]$ is entire on $\bbC^n$, and
\eq{
|\scrC_{\bm A}(\bm z,\rho)|&\le\big\|\tilde f_{\bm A}^{(N,w)}(\bm z)\big\|\,\Tr(\rho)\nonumber\\
&\le\exp\Big(\sum_{j=1}^{n}R_j|\Im z_j|\Big).\label{eq:pw-bound}}
On the real axis $|\scrC_{\bm A}(\bm\theta,\rho)|\le1$ and $\scrC_{\bm A}$ is $C^\infty$.

\emph{Existence, uniqueness, and support.}
By Eq.~\eqref{eq:pw-bound}, $\scrC_{\bm A}(\,\cdot\,,\rho)$ satisfies the hypothesis of Lemma~\ref{lem:pws} with $C=1$, $M=0$.
Therefore there is a unique $\Pr_{\bm A}(\,\cdot\,,\rho)\in\calE'(\bbR^n)$, supported in $\prod_j[-R_j,R_j]$, whose Fourier transform is $\scrC_{\bm A}(\bm\theta,\rho)$, which is Eq.~\eqref{eq:bochner_FT}.
For the product orderings (KD, MH), inserting
\eq{
\e^{\i\theta_jA_j}=\int\e^{\i\theta_jx_j}\d P_{A_j}(x_j)\nonumber
}
shows that the support lies in the joint spectral range $\prod_j\operatorname{spec}(A_j)$, so $\bm x$ is read as the joint measurement outcomes.

\emph{Statement~1 (normalization).}
Since $\Pr_{\bm A}(\,\cdot\,,\rho)\in\calE'(\bbR^n)$ has compact support, it pairs with $1\in C^\infty$, and from Eq.~\eqref{eq:bochner_FT} at $\bm\theta=\bm0$, using $\tilde f_{\bm A}^{(N,w)}(\bm0)=[\sum_\sigma w(\sigma)\openone]^N=\openone$, we get $\langle\Pr_{\bm A}(\cdot,\rho),1\rangle=\scrC_{\bm A}(\bm0,\rho)=\Tr(\rho)=1$.

\emph{Statement~2 (moment generation).}
Compact support implies all moments are finite ($\bm x^\alpha\in C^\infty$).
Differentiating Eq.~\eqref{eq:bochner_FT} under the pairing gives
\eq{
\partial^{|\alpha|}_{\bm\theta}\scrC_{\bm A}(\bm\theta,\rho)=\langle\Pr_{\bm A},(\i\bm x)^\alpha\e^{\i\bm\theta\cdot\bm x}\rangle,\nonumber
}
and at $\bm\theta=\bm0$,
\eq{
\partial^{|\alpha|}_{\bm\theta}\scrC_{\bm A}(\bm\theta,\rho)|_{\bm\theta=\bm0}=\i^{|\alpha|}\langle\Pr_{\bm A},\bm x^\alpha\rangle.\nonumber
}

\emph{Statement~3 (reality).}
A direct computation gives $\widehat{\bar u}(\bm\theta)=\overline{\hat u(-\bm\theta)}$, and $u$ is real iff $u=\bar u$.
Hence $\Pr_{\bm A}$ is real iff $\hat{\Pr}_{\bm A}(\bm\theta)=\overline{\hat{\Pr}_{\bm A}(-\bm\theta)}$, i.e.\ $\scrC_{\bm A}(-\bm\theta,\rho)=\overline{\scrC_{\bm A}(\bm\theta,\rho)}$.

\emph{Statement~4 (non-negativity).}
If $\scrC_{\bm A}(\,\cdot\,,\rho)$ is positive definite, then by Lemma~\ref{lem:bk} there is a Borel probability measure $\mu$ with $\scrC_{\bm A}(\bm\theta,\rho)=\int\exp(\i\bm\theta\cdot\bm x)\d\mu$; by uniqueness $\Pr_{\bm A}(\,\cdot\,,\rho)=\mu$, a compactly supported probability measure.
Conversely, if $\Pr_{\bm A}(\,\cdot\,,\rho)$ is non-negative, then for any $\{\bm t_k\},\{c_k\}$, we have
\eq{
\sum_{j,k}c_j\overline c_k\,\scrC_{\bm A}(\bm t_j-\bm t_k,\rho)&=\int_{\bbR^n}\Big|\sum_k c_k\e^{\i\bm t_k\cdot\bm x}\Big|^2\d\Pr_{\bm A}(\bm x,\rho)\nonumber\\
&\ge0,\nonumber
}
so $\scrC_{\bm A}$ is positive definite.
\end{proofof}

\section{\label{sec:dqsq}Derivations of quantum statistical quantities}
In this appendix, we derive the statistical quantities of the main text, using the canonical purification and the trace operation.
\begin{proofof}{Eq.~\eqref{eq:1st-moment}}
Differentiating the QMGF and using the structure of the purified state $\ket{\Psi}=\sum_i\sqrt{\lambda_i}\ket{\alpha_i}\otimes\bra{\alpha_i}$,
\begin{widetext}
\eq{
\left.\frac{\d}{\d\theta}\scrM_{A}(\theta,\rho)\right|_{\theta=0}&=\left.\frac{\d}{\d\theta}\bra{\Psi}(\e^{\theta A}\otimes\openone_{\calH^{*}})\ket{\Psi}\right|_{\theta=0}\nonumber\\
&=\bra{\Psi}(A\otimes\openone_{\calH^{*}})\ket{\Psi}\nonumber\\
&=\sum_{i,j}\sqrt{\lambda_{i}}\sqrt{\lambda_{j}}\left(\ket{\alpha_{i}}\otimes\bra{\alpha_{i}},\,(A\ket{\alpha_{j}})\otimes\bra{\alpha_{j}}\right)\nonumber\\
&=\sum_{i,j}\sqrt{\lambda_{i}}\sqrt{\lambda_{j}}\,\bra{\alpha_{i}}A\ket{\alpha_{j}}\,(\bra{\alpha_{i}},\bra{\alpha_{j}})\nonumber\\
&=\sum_{i,j}\sqrt{\lambda_{i}}\sqrt{\lambda_{j}}\,\bra{\alpha_{i}}A\ket{\alpha_{j}}\,\delta_{ij}\nonumber\\
&=\sum_{i}\lambda_{i}\bra{\alpha_{i}}A\ket{\alpha_{i}}\nonumber\\
&=\Tr(A\rho)\nonumber\\
&\eqcolon\Ex_{\rho}(A).
}
\end{widetext}
Thus the first derivative reproduces the standard expectation value.
\end{proofof}
\begin{proofof}{Eq.~\eqref{eq:2nd-moment}}
For the centered observable $A_{0} \coloneqq A - \Ex_{\rho}(A)$,
\eq{
\left.\frac{\d^{2}}{\d\theta^{2}}\scrM_{A_{0}}(\theta,\rho)\right|_{\theta=0}&=\left.\bra{\Psi}(A_0^2\,\e^{\theta A_0}\otimes\openone_{\calH^{*}})\ket{\Psi}\right|_{\theta=0}\nonumber\\
&=\bra{\Psi}(A_0^2\otimes\openone_{\calH^{*}})\ket{\Psi}\nonumber\\
&=\Tr\{[A-\Ex_{\rho}(A)]^{2}\rho\}\nonumber\\
&\eqcolon\Var_{\rho}(A).
}
\end{proofof}
\begin{proofof}{Eq.~\eqref{eq:1st-moment_2params}}
For the covariance, we use the MH ordering in the multivariable QMGF.
The mixed partial derivative produces the anticommutator,
\begin{widetext}
\eq{
\left.\frac{\partial^{2}}{\partial\theta_{1}\partial\theta_{2}}\scrM^{\text{MH}}_{A_{0},B_{0}}(\theta_{1},\theta_{2},\rho)\right|_{\theta_{1},\theta_{2}=0}&=\left.\frac{1}{2}\frac{\partial^{2}}{\partial\theta_{1}\partial\theta_{2}}\bra{\Psi}\left[\left(\e^{\theta_{1}A_{0}}\e^{\theta_{2}B_{0}}+\e^{\theta_{2}B_{0}}\e^{\theta_{1}A_{0}}\right)\otimes\openone_{\calH^{*}}\right]\ket{\Psi}\right|_{\theta_{1},\theta_{2}=0}\nonumber\\
&=\frac{1}{2}\bra{\Psi}\left[\left(A_{0}B_{0}+B_{0}A_{0}\right)\otimes\openone_{\calH^{*}}\right]\ket{\Psi}\nonumber\\
&=\frac{1}{2}\Tr[(AB+BA)\rho]-\Tr(A\rho)\Tr(B\rho)\nonumber\\
&\eqcolon\Cov_{\rho}(A,B).
}
\end{widetext}
\end{proofof}
\begin{proofof}{Eq.~\eqref{eq:1st-cond-moment}}
The conditional QMGF is a ratio.
At $\theta=0$, the exponential becomes the identity, and the quotient rule gives
\eq{
\left.\frac{\d}{\d\theta}\scrM_{A}(\theta|\Pi_{m},\rho)\right|_{\theta=0}&=\frac{\bra{\Psi}(\Pi_{m}A\otimes\openone_{\calH^{*}})\ket{\Psi}}{\bra{\Psi}(\Pi_{m}\otimes\openone_{\calH^{*}})\ket{\Psi}}\nonumber\\
&=\frac{\Tr(\Pi_{m}A\rho)}{\Tr(\Pi_{m}\rho)}\nonumber\\
&\eqcolon\Ex_{\rho}(A|\Pi_{m}),
}
the weak value.
\end{proofof}
\begin{proofof}{Eq.~\eqref{eq:2nd-cond-moment}}
For the conditionally centered observable $A'_0\coloneqq A-\Ex_\rho(A|\Pi_m)$, the term linear in the first derivative vanishes, and
\eq{
\left.\frac{\d^{2}}{\d\theta^{2}}\scrM_{A'_{0}}(\theta|\Pi_{m},\rho)\right|_{\theta=0}&=\frac{\bra{\Psi}(\Pi_{m}A_0'^{2}\otimes\openone_{\calH^{*}})\ket{\Psi}}{\bra{\Psi}(\Pi_{m}\otimes\openone_{\calH^{*}})\ket{\Psi}}\nonumber\\
&=\frac{\Tr(\Pi_{m}A^{2}\rho)}{\Tr(\Pi_{m}\rho)}-\left[\frac{\Tr(\Pi_{m}A\rho)}{\Tr(\Pi_{m}\rho)}\right]^{2}\nonumber\\
&=\Ex_{\rho}(A^{2}|\Pi_{m})-\Ex_{\rho}(A|\Pi_{m})^{2}\nonumber\\
&\eqcolon\Var_{\rho}(A|\Pi_{m}).
}
\end{proofof}
\begin{proofof}{Eq.~\eqref{eq:n-point_cor-func2}}
We take the $n$-th mixed derivative of the KD-type QMGF.
Writing $\rho=\sum_i\lambda_i\ketbra{\alpha_i}{\alpha_i}$ and inserting the spectral decompositions $A_j=\sum_{a_j}a_j\ketbra{a_j}{a_j}$ (nondegenerate spectra),
\begin{widetext}
\eq{
\left.\frac{\partial^{n}}{\partial\theta_{1}\cdots\partial\theta_{n}}\scrM^{\text{KD}}_{\bm{A}}(\bm{\theta},\rho)\right|_{\bm\theta=\bm0}
&=\bra{\Psi}(A_{1}A_{2}\cdots A_{n}\otimes\openone_{\calH^{*}})\ket{\Psi}\nonumber\\
&=\sum_{i}\lambda_i\,\bra{\alpha_i}A_1A_2\cdots A_n\ket{\alpha_i}\nonumber\\
&=\sum_{a_{1},\ldots,a_{n},i}a_{1}\cdots a_{n}\,\lambda_{i}\,\braket{\alpha_i}{a_1}\braket{a_1}{a_2}\cdots\braket{a_{n-1}}{a_n}\braket{a_n}{\alpha_i}.\label{eq:np-raw}
}
\end{widetext}
To rewrite this as probabilities and sequential weak values, define, for $k=1,\dots,n-1$, the sequential weak value of Eq.~\eqref{eq:seq-weak},
\eq{
\Ex_{\alpha_i}[P_{A_{k+1}}(a_{k+1})|P_{A_k}(a_k)]&=\frac{\bra{a_k}P_{A_{k+1}}(a_{k+1})\ket{\alpha_i}}{\braket{a_k}{\alpha_i}}\nonumber\\
&=\frac{\braket{a_k}{a_{k+1}}\braket{a_{k+1}}{\alpha_i}}{\braket{a_k}{\alpha_i}},\nonumber
}
and the probability $\Pr\{A_1=a_1\Vert\alpha_i\}=|\braket{a_1}{\alpha_i}|^2=\braket{\alpha_i}{a_1}\braket{a_1}{\alpha_i}$.
Their product telescopes,
\begin{widetext}
\eq{
\Pr\{A_1=a_1\Vert\alpha_i\}\prod_{k=1}^{n-1}\Ex_{\alpha_i}[P_{A_{k+1}}(a_{k+1})|P_{A_k}(a_k)]
&=\braket{\alpha_i}{a_1}\braket{a_1}{\alpha_i}\,\frac{\braket{a_1}{a_2}\cdots\braket{a_{n-1}}{a_n}\,\braket{a_n}{\alpha_i}}{\braket{a_1}{\alpha_i}}\nonumber\\
&=\braket{\alpha_i}{a_1}\braket{a_1}{a_2}\cdots\braket{a_{n-1}}{a_n}\braket{a_n}{\alpha_i},\nonumber
}
\end{widetext}
which is exactly the summand of Eq.~\eqref{eq:np-raw}, where the telescoping uses $\prod_{k=1}^{n-1}\braket{a_{k+1}}{\alpha_i}/\braket{a_k}{\alpha_i}=\braket{a_n}{\alpha_i}/\braket{a_1}{\alpha_i}$.
Substituting back yields Eq.~\eqref{eq:n-point_cor-func2},
\begin{widetext}
\eq{
\left.\frac{\partial^{n}}{\partial\theta_{1}\cdots\partial\theta_{n}}\scrM^{\text{KD}}_{\bm{A}}(\bm{\theta},\rho)\right|_{\bm\theta=\bm0}
=\sum_{a_{1},\ldots,a_{n},i}a_{1}\cdots a_{n}\,\lambda_{i}\,\Pr\{A_{1}=a_{1}\Vert\alpha_{i}\}\prod_{k=1}^{n-1}\Ex_{\alpha_{i}}[P_{A_{k+1}}(a_{k+1})|P_{A_{k}}(a_{k})].\nonumber
}
\end{widetext}
This expresses the $n$-point correlation function as a sequence of weak measurements conditioned on the initial state $\alpha_i$.
\end{proofof}

\section{\label{sec:qgmm}QGMM update and optimality}
In this appendix, we provide a detailed step-by-step derivation of the linearized QGMM update [Eqs.~\eqref{eq:onestep}--\eqref{eq:Jqgmm}] and prove its asymptotic optimality.
We begin with the initial parameter estimate $\bar{\bm\phi}$, the diagonal covariance matrix $\mathbf\Sigma=N^{-1}\openone$, and the Jacobian matrix
\eq{
D=\begin{pmatrix}\beta&0\\0&\beta\\2\beta\hat\mu_1&0\end{pmatrix}.\nonumber
}
To perform the one-step Gauss--Newton update, we first construct the optimal weight matrix $\mathbf{W}_{\text{opt}} = \mathbf\Sigma^{-1} = N\openone$. We then compute the Fisher information-like matrix $D^T\mathbf\Sigma^{-1}D$. Substituting our expressions yields
\eq{
D^T\mathbf\Sigma^{-1}D=N\,\mathrm{diag}(\beta^2(1+4\hat\mu_1^2),\beta^2).
}
Inverting this diagonal matrix gives the asymptotic covariance matrix of the optimal estimator:
\eq{
(D^T\mathbf\Sigma^{-1}D)^{-1}=\mathrm{diag}([\beta^2(1+4\hat\mu_1^2)]^{-1},\beta^{-2})/N.
}
Next, we evaluate the gradient of the objective function. The residual vector, representing the difference between the empirical measurements and the theoretical moments evaluated at the initial estimate, is $\hat{\bm\mu}-\bm\mu(\bar{\bm\phi})=(0,0,\hat\mu_3-\hat\mu_1^2)^T$.
Multiplying this residual by the transposed Jacobian and the optimal weight matrix yields
\eq{
D^T\mathbf\Sigma^{-1}(\hat{\bm\mu}-\bm\mu(\bar{\bm\phi}))=N(2\beta\hat\mu_1(\hat\mu_3-\hat\mu_1^2),0)^T.
}Substituting these components into the generic one-step update formula [Eq.~\eqref{eq:onestep}] exactly produces the explicit analytical estimators $\hat{J}_{\text{QGMM}}$ and $\hat{h}_{\text{QGMM}}$ presented in Eqs.~\eqref{eq:Jqgmm} and \eqref{eq:hqgmm}.
To establish the asymptotic optimality of this estimator, we consider an arbitrary positive definite weight matrix $\mathbf{W}$.
The linearized GMM estimator constructed with such a weight matrix possesses the asymptotic covariance:
\eq{
V(\mathbf{W})=(D^T\mathbf{W}D)^{-1}D^T\mathbf{W}\mathbf\Sigma\mathbf{W}D(D^T\mathbf{W}D)^{-1}.
}
According to the generalized Aitken (or Gauss--Markov) theorem \cite{Hansen82}, the matrix inequality $V(\mathbf{W})\succeq(D^T\mathbf\Sigma^{-1}D)^{-1}=V(\mathbf\Sigma^{-1})$ holds for all valid weighting matrices $\mathbf{W}$, with equality strictly achieved when $\mathbf{W}=\mathbf\Sigma^{-1}$.
Therefore, setting the weight matrix to the inverse of the quantum covariance matrix explicitly minimizes the estimator's variance.
Specifically, the minimal variance for the parameter $J$ is given by the $(1,1)$ entry of the optimal covariance matrix $(D^T\mathbf\Sigma^{-1}D)^{-1}$, which is exactly $1/[N\beta^2(1+4\hat\mu_1^2)]\le1/(N\beta^2)$.
Because this value is strictly less than or equal to the QMM variance $1/(N\beta^2)$, the QGMM estimator is demonstrably optimal, providing a quantifiable efficiency gain whenever over-identifying conditions are incorporated.

\section{\label{sec:dgmbf}Derivations for alternative approaches}
We derive the statistical properties of the QMGF defined via the matrix geometric mean, using the variational properties of the geometric mean and perturbation theory for the matrix square root.
\begin{proofof}{Eq.~\eqref{eq:1st-moment_gm}}
The key step is the derivative of the matrix square root, which solves a Lyapunov (Sylvester) equation.
Let $X(\theta) = \rho^{-1/2} \exp(\theta V) \rho^{-1/2}$.
For small $\theta$, $\exp(\theta V) \approx \openone + \theta V$, so
\eq{
X(\theta) \approx \rho^{-1} + \theta\, \rho^{-1/2} V \rho^{-1/2}.
}
With $\scrM^{\text{geo}}_{V}(\theta, \rho)=\Tr[\rho^{1/2} \sqrt{X(\theta)} \rho^{1/2}]$ and $Y \coloneqq \left. \d_\theta \sqrt{X(\theta)} \right|_{\theta=0}$,
\eq{
\left. \frac{\d}{\d\theta} \scrM^{\text{geo}}_{V}(\theta, \rho) \right|_{\theta=0} = \Tr\left( \rho^{1/2} Y \rho^{1/2} \right).
}
Differentiating $(\sqrt{X})^2=X$ at $\theta=0$ gives the Lyapunov equation $\rho^{-1/2} Y + Y \rho^{-1/2} = \rho^{-1/2} V \rho^{-1/2}$.
In the eigenbasis $\rho = \sum_k p_k \ketbra{k}{k}$,
\eq{
(p_k^{-1/2} + p_l^{-1/2})\,Y_{kl} = \frac{V_{kl}}{\sqrt{p_k p_l}}
}
gives
\eq{
Y_{kl} = \frac{V_{kl}}{\sqrt{p_k} + \sqrt{p_l}}.
}
Therefore
\eq{
\Tr(\rho Y) &= \sum_k p_k Y_{kk}\nonumber\\
&=\sum_k p_k \frac{V_{kk}}{2\sqrt{p_k}}\nonumber\\
&= \frac{1}{2} \sum_k \sqrt{p_k}\,V_{kk}\nonumber\\
&= \frac{1}{2}\Tr(\sqrt{\rho} V).
}
The first derivative is thus the $\sqrt{\rho}$-weighted expectation value, reducing to $\Tr(\rho V)$ only when $[\rho, V] = 0$.
\end{proofof}
\begin{proofof}{Eq.~\eqref{eq:2nd-moment_gm}}
Expanding $X(\theta) = \rho^{-1/2} \exp(\theta V) \rho^{-1/2}$ to second order,
\eq{
X(\theta) \approx \rho^{-1} + \theta\, \rho^{-1/2} V \rho^{-1/2} + \frac{\theta^2}{2}\, \rho^{-1/2} V^2 \rho^{-1/2},
}
and writing $\sqrt{X(\theta)} \approx Y_0 + \theta Y_1 + \tfrac{\theta^2}{2} Y_2$ with $Y_0 = \rho^{-1/2}$ and $(Y_1)_{kl} = V_{kl} / (\sqrt{p_k} + \sqrt{p_l})$, the order-$\theta^2$ terms of $(\sqrt X)^2=X$ give
\eq{
\rho^{-1/2} Y_2 + Y_2 \rho^{-1/2} = \rho^{-1/2} V^2 \rho^{-1/2} - 2 Y_1^2.
}
The diagonal elements are
\eq{
(Y_2)_{kk} = \frac{(V^2)_{kk}}{2\sqrt{p_k}} - \sqrt{p_k}\,(Y_1^2)_{kk}.
}
The second derivative is
\begin{widetext}
\eq{
\left. \frac{\d^2 \scrM^{\text{geo}}_{V}(\theta, \rho)}{\d \theta^2} \right|_{\theta=0} &= \Tr(\rho Y_2) \nonumber\\
&= \sum_k p_k \left[ \frac{(V^2)_{kk}}{2\sqrt{p_k}} - \sqrt{p_k} \sum_j |(Y_1)_{kj}|^2 \right] \nonumber\\
&= \frac{1}{2}\Tr(\sqrt{\rho} V^2) - \sum_{k,j} \frac{p_k^{3/2}}{(\sqrt{p_k} + \sqrt{p_j})^2} |\bra{k}V\ket{j}|^2.
}
\end{widetext}
\end{proofof}
\begin{proofof}{Eq.~\eqref{eq:1st-moment_is}}
Let $f(\theta) = \Tr(\rho^{1-\theta} \sigma^{\theta})$, so $\psi'(\theta) = f'(\theta)/f(\theta)$ and, since $f(0)=\Tr\rho=1$, $\psi'(0)=f'(0)$.
Using $\rho^{1-\theta} \approx \rho (\openone - \theta \log \rho)$ and $\sigma^{\theta} \approx \openone + \theta \log \sigma$,
\eq{
f(\theta) &\approx \Tr\left[ \rho + \theta \rho (\log \sigma - \log \rho) \right] + O(\theta^2),
}
so $f'(0) = \Tr[\rho (\log \sigma - \log \rho)] = -D(\rho\|\sigma)$, giving $\psi'(0)=-D(\rho\|\sigma)$.
\end{proofof}
\begin{proofof}{Eq.~\eqref{eq:2nd-moment_is}}
From $\psi'(\theta)=f'(\theta)/f(\theta)$ and $f(0)=1$, $\psi''(0)=f''(0)-[f'(0)]^2$.
With $A=\log\rho$, $B=\log\sigma$, expanding $\rho^{1-\theta} \approx \rho - \theta \rho A + \tfrac{\theta^2}{2} \rho A^2$ and $\sigma^\theta \approx \openone + \theta B + \tfrac{\theta^2}{2} B^2$,
\eq{
f(\theta) &\approx \Tr(\rho) + \theta \Tr[\rho(B-A)] \nonumber\\
&\mbox{}\quad+ \theta^2 \left[ \tfrac{1}{2}\Tr(\rho A^2) + \tfrac{1}{2}\Tr(\rho B^2) - \Tr(\rho A B) \right] \nonumber\\
&\mbox{}\quad+ O(\theta^3),
}
so that, using $[\rho,A]=0$,
\eq{
f''(0) &= \Tr[\rho (A^2 - 2AB + B^2)] \nonumber\\
&= \Tr[\rho (B-A)^2] \nonumber\\
&=\Tr[\rho(\log\sigma-\log\rho)^2].
}
Therefore
\eq{
\psi''(0) &= \Tr[ \rho (\log \sigma - \log \rho)^2 ] - ( \Tr[\rho (\log \sigma - \log \rho)] )^2 \nonumber\\
&\eqcolon \Var(\rho\|\sigma),
}
the quantum relative entropy variance.
\end{proofof}

\sloppy

\bibliography{ref}

\end{document}